\documentclass[fleqn,prb,aps,eqsecnum,amsmath,amssymb,floatfix]{revtex4}
\usepackage[T1]{fontenc}
\usepackage[latin9]{inputenc}
\usepackage[active]{srcltx}
\usepackage{amstext}
\usepackage{graphicx}

\makeatletter
\@ifundefined{textcolor}{}
{%
 \definecolor{BLACK}{gray}{0}
 \definecolor{WHITE}{gray}{1}
 \definecolor{RED}{rgb}{1,0,0}
 \definecolor{GREEN}{rgb}{0,1,0}
 \definecolor{BLUE}{rgb}{0,0,1}
 \definecolor{CYAN}{cmyk}{1,0,0,0}
 \definecolor{MAGENTA}{cmyk}{0,1,0,0}
 \definecolor{YELLOW}{cmyk}{0,0,1,0}
}

\makeatother

\begin{document}

\title{Composite Operator Method analysis of the underdoped cuprates puzzle}

\author{Adolfo Avella}

\affiliation{Dipartimento di Fisica ``E.R. Caianiello'', Universit\`a degli
Studi di Salerno, I-84084 Fisciano (SA), Italy}

\affiliation{Unit\`a CNISM di Salerno, Universit\`a degli Studi di Salerno,
I-84084 Fisciano (SA), Italy}

\affiliation{CNR-SPIN, UoS di Salerno, I-84084 Fisciano (SA), Italy}

\begin{abstract}
The microscopical analysis of the unconventional and puzzling physics
of the underdoped cuprates, as carried out lately by means of the
Composite Operator Method (COM) applied to the two-dimensional (2D)
Hubbard model, is reviewed and systematized. The 2D Hubbard model
has been adopted as, since the very early days of unconventional high-$T_{c}$
superconductivity, it has been considered the minimal model capable
to describe the most peculiar features of cuprates held responsible
for their anomalous behavior. As a matter of fact, understanding the
physics of the 2D Hubbard model itself constitutes one of the most
intriguing challenges in condensed matter theory. In the last fifteen
years, COM has proved to be a quite powerful non-perturbative, fully-analytical,
self-consistent, microscopical approximation methods specifically
devised to deal with strongly correlated systems (SCSs). COM is designed
to endorse, since its foundations, the systematic emergence in any
SCS of new elementary excitations described by composite operators
obeying non-canonical algebras. COM is formulated to deal with the
unusual features of such composite operators and compute the unconventional
properties of SCSs. In this case (underdoped cuprates -- 2D Hubbard
model), the residual interactions -- beyond a $2$-pole approximation
-- between the new elementary electronic excitations, dictated by
the strong local Coulomb repulsion and well described by the two Hubbard
composite operators, have been treated within the Non Crossing Approximation
(NCA). The two-particle spin and charge propagators, appearing in
the electronic self-energy thanks to the composite nature of the new
elementary electronic excitations, have been computed fully-microscopically
within the very same framework, just neglecting any explicit damping
in a first approximation. Given this recipe and exploiting the few
unknowns to enforce the Pauli principle content in the solution, it
is possible to qualitatively describe -- finite and specific longer-distance
hopping terms are needed for a quantitative comparison to a specific
material -- some of the anomalous features of high-T$_{c}$ cuprate
superconductors such as large vs. small Fermi surface dichotomy, Fermi
surface deconstruction (appearance of Fermi arcs), nodal vs. anti-nodal
physics, pseudogap(s), kinks in the electronic dispersion. The resulting
scenario envisages a smooth crossover between an ordinary weakly-interacting
metal sustaining weak, short-range antiferromagnetic correlations
in the overdoped regime to an unconventional poor metal characterized
by very strong, long-but-finite-range antiferromagnetic correlations
leading to momentum-selective non-Fermi liquid features as well as
to the opening of a pseudogap and to the striking differences between
the nodal and the anti-nodal dynamics in the underdoped regime.
\end{abstract}


\maketitle

\section{Introduction}

\subsection{Composite fields}

One of the most intriguing challenges in modern condensed matter physics
is the theoretical description of the anomalous behaviors experimentally
observed in many novel materials. By anomalous behaviors we mean those
not predicted by standard many-body theory; that is, behaviors in
contradiction with the Fermi-liquid framework and diagrammatic expansions.
The most relevant characteristic of such novel materials is the presence
of so strong correlations among the electrons that classical schemes
based on the band picture and the perturbation theory are definitely
inapplicable. Accordingly, it is necessary to move from a \emph{single-electron}
physics to a \emph{many-electron} physics, where the dominant contributions
come from the strong interactions among the electrons: usual schemes
are simply inadequate and new concepts must be introduced.

The \emph{classical} techniques are based on the hypothesis that the
interactions among the electrons are weak enough, or sufficiently
well screened, to be properly taken into account within the framework
of perturbative/diagrammatic methods. However, as many and many experimental
and theoretical studies of highly correlated systems have shown, with
more and more convincing evidence, all these methods are no more viable.
The main concept that breaks down is the existence of the electrons
as particles or quasi-particles with quite-well-defined properties.
The presence of the interactions radically modifies the properties
of the particles and, at a macroscopic level, what are observed are
new particles (actually they are the only observable ones) with new
peculiar properties entirely determined by the dynamics and by the
boundary conditions (i.e. the phase under study, the external fields,
...). These new objects appear as the final result of the modifications
imposed by the interactions on the original particles and contain,
by the very beginning, the effects of correlations.

On the basis of this evidence, one is induced to move the attention
from the original fields to the new fields generated by the interactions.
The operators describing these excitations, once they have been identified,
can be written in terms of the original ones and are known as composite
operators. The necessity of developing a formulation to treat composite
operators as fundamental objects of the many-body problem in condensed
matter physics has been deeply understood and systematically noticed
since quite long time. Recent years have seen remarkable achievements
in the development of a modern many-body theory in solid-state physics
in the form of an assortment of techniques that may be termed composite
particle methods. The foundations of these types of techniques may
be traced back to the work of Bogoliubov \cite{Bogoliubov_47} and
later to that of Dancoff \cite{Dancoff_50}. The work of Zwanzig \cite{Zwanzig_61},
Mori \cite{Mori_65,Plakida_95,Plakida_01,Plakida_06,Adam_07,Plakida_10}
and Umezawa \cite{Umezawa_93} definitely deserves to be mentioned
too. Closely related to this work is that of Hubbard \cite{Hubbard_63,Hubbard_64,Hubbard_64a},
Rowe \cite{Rowe_68}, Roth \cite{Roth_69} and Tserkovnikov \cite{Tserkovnikov_81,Tserkovnikov_81a}.
The slave boson method \cite{Barnes_76,Coleman_84,Kotliar_86}, the
spectral density approach \cite{Kalashnikov_69,Nolting_72}, the diagram
technique for Hubbard operators \cite{Izyumov_89}, the cumulant expansion
based diagram technique \cite{Moskalenko_90}, the generalized tight-binding
method \cite{Ovchinnikov_04,Korshunov_05,Ovchinnikov_06}, self-consistent
projection operator method \cite{Kakehashi_04,Kakehashi_05}, operator
projection method \cite{Onoda_01,Onoda_01a,Onoda_03} and the composite
operator method (COM) \cite{Theory,Avella_11a} are along the same
lines. This large class of theories is very promising as it is based
on the firm conviction that strong interactions call for an analysis
in terms of new elementary fields embedding the greatest possible
part of the correlations so permitting to overcome the problem of
finding an appropriate expansion parameter. However, one price must
be paid. In general, composite fields are neither Fermi nor Bose operators,
since they do not satisfy canonical (anti)commutation relations, and
their properties must be determined self-consistently. They can only
be recognized as fermionic or bosonic operators according to the number
and type of the constituting original particles. Accordingly, new
techniques have to be developed in order to deal with such composite
fields and to design diagrammatic schemes where the building blocks
are the propagators of such composite fields: standard diagrammatic
expansions and the Wick's theorem are no more valid. The formulation
of the Green's function method itself must be revisited and new frameworks
of calculations have to be devised.

Following these ideas, we have been developing a systematic approach,
the composite operator method (COM) \cite{Theory,Avella_11a}, to
study highly correlated systems. The formalism is based on two main
ideas: (i) use of propagators of relevant composite operators as building
blocks for any subsequent approximate calculations; (ii) use of algebra
constraints to fix the representation of the relevant propagators
in order to properly preserve algebraic and symmetry properties; these
constraints will also determine the unknown parameters appearing in
the formulation due to the non-canonical algebra satisfied by the
composite operators. In the last fifteen years, COM has been applied
to several models and materials: Hubbard \cite{Hub,Avella_03c,Krivenko_04,Odashima_05,Avella_14},
$p$-$d$ \cite{p-d}, $t$-$J$ \cite{Avella_02a}, $t$-$t'$-$U$
\cite{ttU}, extended Hubbard ($t$-$U$-$V$) \cite{tUV}, Kondo
\cite{Villani_00}, Anderson \cite{Anderson}, two-orbital Hubbard
\cite{2orb,Plekhanov_11}, Ising \cite{Ising}, $J_{1}-J_{2}$ \cite{Bak_02a,J1J2,Avella_08a},
Hubbard-Kondo \cite{Avella_06a}, Cuprates \cite{Cuprates-NCA,Avella_07,Avella_07a,Avella_08,Avella_09},
etc.

\subsection{Underdoped cuprates\label{sec:Materials}}

Cuprate superconductors \cite{Bednorz_86} display a full range of
anomalous features, mainly appearing in the underdoped region, in
almost all experimentally measurable physical properties \cite{Timusk_99,Orenstein_00,Damascelli_03,Eschrig_06,Sebastian_12a}.
According to this, their microscopic description is still an open
problem: non-Fermi-liquid response, quantum criticality, pseudogap
formation, ill-defined Fermi surface, kinks in the electronic dispersion,
etc. remain still unexplained (or at least controversially debated)
anomalous features \cite{Lee_06,Tremblay_06,Sebastian_12a}. In the
last years, the attention of the community has been focusing on three
main experimental facts \cite{Sebastian_12a}: the dramatic change
in shape and nature of the Fermi surface between underdoped and overdoped
regimes, the appearance of a psedudogap in the underdoped regime,
and the striking differentiation between the physics at the nodes
and at the anti-nodes in the pseudogap regime. The topological transition
of the Fermi surface has been first detected by means of ARPES \cite{Damascelli_03}
and reflects the noteworthy differences between the quite-ordinary,
large Fermi surface measured in the overdoped regime \cite{Hussey_03,Plate_05,Peets_07}
and quite well described by LDA calculations \cite{Andersen_95} and
quantum oscillations measurements \cite{Vignolle_08}, and the ill-defined
Fermi arcs appearing in the underdoped regime \cite{Shen_05,Valla_06,Lee_06,Kanigel_06,Hossain_08,Meng_09,King_11}.
The enormous relevance of these experimental findings, not only for
the microscopic comprehension of the high-$T_{c}$ superconductivity
phenomenology, but also for the drafting of a general microscopic
theory for strongly correlated materials, called for many more measurements
in order to explore all possible aspects of such extremely anomalous
and peculiar behavior: plenty of quantum oscillations measurements
in the underdoped regime \cite{Doiron-Leyraud_07,LeBoeuf_07,Yelland_08,Bangura_08,Sebastian_08,Audouard_09,Sebastian_10,Sebastian_10a,Sebastian_10b,Singleton_10,Vishik_10,Anzai_10,Ramshaw_11,Sebastian_11,Sebastian_11a,Sebastian_11b,Riggs_11,Laliberte_11,Vignolle_11,Sebastian_12},
Hall effect measurements \cite{LeBoeuf_07}, Seebeck effect measurements
\cite{Laliberte_11}, and heat capacity measurements \cite{Riggs_11}.
The presence of a quite strong depletion in the electronic density
of states, known as pseudogap \cite{Norman_05}, is well established
thanks to ARPES \cite{Damascelli_03}, NMR \cite{Alloul_89}, optical
conductivity \cite{Basov_99} and quantum oscillations \cite{Tranquada_10}
measurements. The microscopic origin of such a loss of single-particle
electronic states is still unclear and the number of possible theoretical,
as well as phenomenological, explanations has grown quite large in
the last few years. As a matter of fact, this phenomenon affects any
measurable properties and, accordingly, was the first to be detected
in the underdoped regime granting to this latter the first evidences
of its exceptionality with respect to the other regimes in the phase
diagram. The plethora of theoretical scenarios present in the literature
\cite{Kivelson_03,Lee_06,Carrington_07,Sushkov_11}, tentatively explaining
few, some or many of the anomalous features reported by the experiments
on underdoped cuprates, can be coarsely divided between those not
relying on any translational symmetry breaking \cite{Alexandrov_96,Yang_06,Melikyan_08,Varma_09,Wilson_09,Pereg-Barnea_10}
and those instead proposing that it should be some kind of charge
and/or spin arrangements to be held responsible for the whole range
of anomalous features. Among these latter theories, there are those
focusing on the physics at the anti-nodal region and those focusing
on the nodal region. We can account for proposals of (as regards the
anti-node): a collinear spin AF order \cite{Millis_07}, an AF quantum
critical point \cite{Haug_10,Baledent_11}, a 1D charge stripe order
\cite{Millis_07,Harrison_11a} with the addition of a smectic phase
\cite{Yao_11}. Instead, at the node, we have: a $d$-density wave
\cite{Chakravarty_08}, a more-or-less ordinary AF spin order \cite{Chubukov_97,Avella_07,Avella_07a,Avella_07b,Avella_07c,Chen_08,Avella_08,Avella_09,Oh_11},
a nodal pocket from bilayer low-$Q$ charge order, slowly fluctuacting
\cite{Li_06,Harrison_11,Harrison_11a,Harrison_12}. This latter proposal,
which is among the newest on the table, relies on many experimental
measurements: STM \cite{Edwards_94,Hoffman_02,Hanaguri_04,McElroy_05,Maki_05},
Neutron scattering \cite{Tranquada_10}, X-ray diffraction \cite{Liu_08},
NMR \cite{Wu_11}, RXS \cite{Ghiringhelli_12}, phonon softening \cite{Mook_96,Mook_99,Reznik_08}.
A proposal regarding the emergence of a hidden Fermi liquid \cite{Anderson_09}
is also worth mentioning.

\subsection{2D Hubbard Model -- Approximation Methods}

Since the very beginning \cite{Anderson_87}, the two-dimensional
Hubbard model \cite{Hubbard_63} has been universally recognized as
the minimal model capable to describe the $Cu-O_{2}$ planes of cuprates
superconductors. It certainly contains many of the key ingredients
by construction: strong electronic correlations, competition between
localization and itineracy, Mott physics, and low-energy spin excitations.
Unfortunately, although fundamental for benchmarking and fine tuning
analytical theories, numerical approaches \cite{Bulut_02} cannot
be of help to solve the puzzle of underdoped cuprates owing to their
limited resolution in frequency and momentum. On the other hand, there
are not so many analytical approaches capable to deal with the quite
complex aspects of underdoped cuprates phenomenology \cite{Plakida_10}.
Among others, the Two-Particle Self-Consistent (TPSC) approach \cite{Vilk_95,Tremblay_06}
has been the first completely microscopic approach to obtain results
comparable with the experimental findings. Almost all other promising
approaches available in the literature can be essentially divided
into two classes. One class makes use of phenomenological expressions
for the electronic self-energy and the electronic spin susceptibility
\cite{Plakida_95,Plakida_01,Chubukov_04,Prelovsek_05,Plakida_06,Adam_07,Plakida_10}.
The electronic self-energy is usually computed as the convolution
of the electronic propagator and of the electronic spin susceptibility.
Then, the electronic spin susceptibility is modeled phenomenologically
parameterizing correlation length and damping as functions of doping
and temperature according to the common belief that the electronic
spin susceptibility should present a well developed mode at $M=(\pi,\pi)$
with a damping of Landau type. The DMFT$+\Sigma$ approach \cite{Sadovskii_01,Sadovskii_05,Kuchinskii_05,Kuchinskii_06,Kuchinskii_06a}
also belongs to this class. All cluster-dynamical-mean-field-like
theories (cluster-DMFT theories) \cite{Maier_05,Tremblay_06,Stanescu_06}
(the cellular dynamical mean-field theory (C-DMFT) \cite{Kotliar_01a,Haule_07},
the dynamical cluster approximation (DCA) \cite{Hettler_98} and the
cluster perturbation theory (CPT) \cite{Senechal_00,Senechal_04,Tremblay_06})
belong to the second class. The dynamical Mean-Field Theory (DMFT)
\cite{Georges_96,Kotliar_04,Kotliar_06,Held_07,Izyumov_08} cannot
tackle the underdoped cuprates puzzle because its self-energy has
the same identical value at each point on the Fermi surface without
any possible differentiation between nodal and antinodal physics or
visible and phantom portion of the Fermi surface. The cluster-DMFT
theories instead can, in principle, deal with both coherent quasi-particles
and marginal ones within the same Fermi surface. These theories usually
self-consistently map the generic Hubbard problem to a few-site lattice
Anderson problem and solve this latter by means of, mainly, numerical
techniques. What really distinguishes one formulation from another,
within this second class, is the procedure used to map the small cluster
on the infinite lattice. Anyway, it is worth noticing that these approaches
often relies on numerical methods in order to close their self-consistency
cycles (with the above mentioned limitations in frequency and momentum
resolutions and with the obvious difficulties in the physical interpretation
of their results) and always face the emergence of a quite serious
periodization problem since a cluster embedded in the lattice violates
its periodicity. The Composite Operator Method (COM) \cite{Theory,Avella_11a}
does not belong to any of these two classes of theoretical formulations
and has the advantage to be completely microscopic, exclusively analytical,
and fully self-consistent. COM recipe uses three main ingredients
\cite{Theory,Avella_11a}: \emph{composite} operators, \emph{algebra}
constraints, and \emph{residual} self-energy treatment. Composite
operators are products of electronic operators and describe the new
elementary excitations appearing in the system owing to strong correlations.
According to the system under analysis \cite{Theory,Avella_11a},
one has to choose a set of composite operators as operatorial basis
and rewrite the electronic operators and the electronic Green's function
in terms of this basis. One should think of composite operators just
as a more convenient starting point, with respect to electronic operators,
for any mean-field-like approximation/perturbation scheme. Algebra
constraints are relations among correlation functions dictated by
the non-canonical operatorial algebra closed by the chosen operatorial
basis \cite{Theory,Avella_11a}. Other ways to obtain algebra constraints
rely on the symmetries enjoined by the Hamiltonian under study, the
Ward-Takahashi identities, the hydrodynamics, etc \cite{Theory,Avella_11a}.
One should think of algebra constraints as a way to restrict the Fock
space on which the chosen operatorial basis acts to the Fock space
of physical electrons. Algebra constraints are used to compute unknown
correlation functions appearing in the calculations. Interactions
among the elements of the chosen operatorial basis are described by
the residual self-energy, that is, the propagator of the residual
term of the current after this latter has been projected on the chosen
operatorial basis \cite{Theory,Avella_11a}. According to the physical
properties under analysis and the range of temperatures, dopings,
and interactions you want to explore, one has to choose an approximation
to compute the residual self-energy. In the last years, we have been
using the $n-$pole Approximation \cite{Hub,Odashima_05,p-d,Avella_02a,ttU,tUV,2orb,Plekhanov_11,Ising,Avella_06a,Cuprates-NCA},
the Asymptotic Field Approach \cite{Villani_00,Anderson}, the NCA
\cite{Avella_03h,Avella_03c,Krivenko_04,Avella_07,Avella_07a,Avella_08,Avella_09}
and the Two-Site Resolvent Approach \cite{Matsumoto_96,Matsumoto_97}.
You should think of the residual self-energy as a measure in the frequency
and momentum space of how much well defined are, as quasi-particles,
your composite operators. It is really worth noticing that, although
the description of some of the anomalous features of underdoped cuprates
given by COM qualitatively coincide with those obtained by TPSC \cite{Tremblay_06}
and by the two classes of formulations mentioned above, the results
obtained by means of COM greatly differs from those obtained within
the other methods as regards the evolution with doping of the dispersion
and of the Fermi surface and, at the moment, no experimental result
can tell which is the unique and distinctive choice nature made.

\subsection{Outline}

To study the underdoped cuprates modeled by the 2D Hubbard model (see
Sec.~\ref{sec:Hamiltonian}), we start from a basis of two composite
operators (the two Hubbard operators) and formulate the Dyson equation
(see Sec.~\ref{sec2.1}) in terms of the $2$-pole approximated Green's
function (see Sec.~\ref{sec:Two-pole-Approximation}). According
to this, the self-energy is the propagator of non-local composite
operators describing the electronic field dressed by charge, spin,
and pair fluctuations on the nearest-neighbor sites. Then, within
the Non-Crossing Approximation (NCA) \cite{Bosse_78}, we obtain a
\emph{microscopic} self-energy written in terms of the convolution
of the electronic propagator and of the charge, spin, and pair susceptibilities
(see Sec.~\ref{s3.3}). Finally, we close, fully analytically, the
self-consistency cycle for the electronic propagator by computing
\emph{microscopic} susceptibilities within a $2$-pole approximation
(see Sec.~\ref{s3.4}). Our results (see Sec.~\ref{s3.3}) show
that, within COM, the two-dimensional Hubbard model can describe some
of the anomalous features experimentally observed in underdoped cuprates
phenomenology. In particular, we show how Fermi arcs can develop out
of a large Fermi surface (see Sec.~\ref{sec:Spectral-Function-and}),
how pseudogap can show itself in the dispersion (see Sec.~\ref{sec:Dispersion})
and in the density of states (see Sec.~\ref{sec:Density-of-States}),
how non-Fermi liquid features can become apparent in the momentum
distribution function (see Sec.~\ref{sec:Momentum-Distribution-Function})
and in the frequency and temperature dependences of the self-energy
(see Sec.~\ref{sec:Self-energy}), how much \emph{kinked} the dispersion
can get on varying doping (see Sec.~\ref{sec:Dispersion}), and why,
or at least how, spin-dynamics can be held responsible for all this
(see Sec.~\ref{sec:Spin-dynamics}). Finally (see Sec.~\ref{sec:Conclusions}),
we summarize the current status of the scenario emerging by these
theoretical findings and which are the perspectives.

\section{Framework}

\subsection{Hamiltonian\label{sec:Hamiltonian}}

The Hamiltonian of the two-dimensional Hubbard model reads as 
\begin{equation}
H=\sum_{\mathbf{ij}}\left(-\mu\delta_{\mathbf{ij}}-4t\alpha_{\mathbf{ij}}\right)c^{\dagger}(i)c(j)+U\sum_{\mathbf{i}}n_{\uparrow}(i)n_{\downarrow}(i)\label{3.1}
\end{equation}
where 
\begin{equation}
c(i)=\left(\begin{array}{c}
c_{\uparrow}(i)\\
c_{\downarrow}(i)
\end{array}\right)\label{3.2}
\end{equation}
is the electron field operator in spinorial notation and Heisenberg
picture ($i=(\mathbf{i},t_{i})$), $\mathbf{i}=\mathbf{R_{i}}$ is
a vector of the Bravais lattice, $n_{\sigma}(i)=c_{\sigma}^{\dagger}(i)c_{\sigma}(i)$
is the particle density operator for spin $\sigma$, $n(i)=\sum_{\sigma}n_{\sigma}(i)$
is the total particle density operator, $\mu$ is the chemical potential,
$t$ is the hopping integral and the energy unit, $U$ is the Coulomb
on-site repulsion and $\alpha_{\mathbf{ij}}$ is the projector on
the nearest-neighbor sites 
\begin{eqnarray}
\alpha_{\mathbf{ij}} & = & \frac{1}{N}\sum_{\mathbf{k}}\mathrm{e}^{\mathrm{i}\mathbf{k}\cdot(\mathbf{R_{i}}-\mathbf{R_{j}})}\alpha(\mathbf{k})\nonumber \\
\alpha(\mathbf{k}) & = & \frac{1}{2}\left[\cos(k_{x}a)+\cos(k_{y}a)\right]\label{3.3}
\end{eqnarray}
where $\mathbf{k}$ runs over the first Brillouin zone, $N$ is the
number of sites and $a$ is the lattice constant.

\subsection{Green's functions and Dyson equation\label{sec2.1}}

Following COM prescriptions \cite{Theory,Avella_11a}, we chose a
basic field; in particular, we select the composite doublet field
operator 
\begin{equation}
\psi(i)=\left(\begin{array}{c}
\xi(i)\\
\eta(i)
\end{array}\right)\label{3.4}
\end{equation}
where $\eta(i)=n(i)c(i)$ and $\xi(i)=c(i)-\eta(i)$ are the Hubbard
operators describing the main subbands. This choice is guided by the
hierarchy of the equations of motion and by the fact that $\xi(i)$
and $\eta(i)$ are eigenoperators of the interacting term in the Hamiltonian
(\ref{3.1}). The field $\psi(i)$ satisfies the Heisenberg equation
\begin{equation}
\mathrm{i}\frac{\partial}{\partial t}\psi(i)=J(i)=\left(\begin{array}{c}
-\mu\xi(i)-4tc^{\alpha}(i)-4t\pi(i)\\
(U-\mu)\eta(i)+4t\pi(i)
\end{array}\right)\label{3.5}
\end{equation}
where the higher-order composite field $\pi(i)$ is defined by 
\begin{equation}
\pi(i)=\frac{1}{2}\sigma^{\mu}n_{\mu}(i)c^{\alpha}(i)+c(i)c^{\dagger\alpha}(i)c(i)\label{3.6}
\end{equation}
with the following notation: $n_{\mu}(i)=c^{\dagger}(i)\sigma_{\mu}c(i)$
is the particle- ($\mu=0$) and spin- ($\mu=1,\,2,\,3$) density operator,
$\sigma_{\mu}=\left(1,\,\vec{\sigma}\right)$, $\sigma^{\mu}=\left(-1,\,\vec{\sigma}\right)$,
$\sigma_{k}\,\left(k=1,\,2,\,3\right)$ are the Pauli matrices. Hereafter,
for any operator $\Phi(i)$, we use the notation $\Phi^{\alpha}(\mathbf{i},t)=\sum_{\mathbf{j}}\alpha_{\mathbf{ij}}\Phi(\mathbf{j},t)$.

It is always possible to decompose the source $J(i)$ under the form
\begin{equation}
J(i)=\varepsilon(-\mathrm{i}\nabla)\psi(i)+\delta J(i)\label{3.7}
\end{equation}
where the linear term represents the projection of the source on the
basis $\psi(i)$ and is calculated by means of the equation 
\begin{equation}
\langle\{\delta J(\mathbf{i},t),\,\psi^{\dagger}(\mathbf{j},t)\}\rangle=0\label{3.8}
\end{equation}
where $\langle\cdots\rangle$ stands for the thermal average taken
in the grand-canonical ensemble.

This constraint assures that the residual current $\delta J(i)$ contains
all and only the physics orthogonal to the chosen basis $\psi(i)$.
The action of the derivative operator $\varepsilon(-\mathrm{i}\nabla)$
on $\psi(i)$ is defined in momentum space 
\begin{eqnarray*}
\varepsilon(-\mathrm{i}\nabla)\psi(i) & = & \varepsilon(-\mathrm{i}\nabla)\frac{1}{\sqrt{N}}\sum_{\mathbf{k}}\mathrm{e}^{\mathrm{i}\mathbf{k}\cdot\mathbf{R_{i}}}\psi(\mathbf{k},t)=\frac{1}{\sqrt{N}}\sum_{\mathbf{k}}\mathrm{e}^{\mathrm{i}\mathbf{k}\cdot\mathbf{R_{i}}}\varepsilon(\mathbf{k})\psi(\mathbf{k},t)
\end{eqnarray*}
where $\varepsilon(\mathbf{k})$ is named energy matrix.

The constraint (\ref{3.8}) gives 
\begin{equation}
m(\mathbf{k})=\varepsilon(\mathbf{k})I(\mathbf{k})\label{3.10}
\end{equation}
after defining the normalization matrix 
\begin{equation}
I(\mathbf{i,j})=\langle\{\psi(\mathbf{i},t),\,\psi^{\dagger}(\mathbf{j},t)\}\rangle=\frac{1}{N}\sum_{\mathbf{k}}\mathrm{e}^{\mathrm{i}\mathbf{k}\cdot(\mathbf{R_{i}-R_{j}})}I(\mathbf{k})\label{3.11}
\end{equation}
and the $m$-matrix 
\begin{equation}
m(\mathbf{i,j})=\langle\{J(\mathbf{i},t),\,\psi^{\dagger}(\mathbf{j},t)\}\rangle=\frac{1}{N}\sum_{\mathbf{k}}\mathrm{e}^{\mathrm{i}\mathbf{k}\cdot(\mathbf{R_{i}-R_{j}})}m(\mathbf{k})\label{3.12}
\end{equation}

Since the components of $\psi(i)$ contain composite operators, the
normalization matrix $I(\mathbf{k})$ is not the identity matrix and
defines the spectral content of the excitations. In fact, the composite
operator method has the advantage of describing crossover phenomena
as the phenomena in which the weight of some operator is shifted to
another one.

By considering the two-time thermodynamic Green's functions \cite{Bogoliubov_59,Zubarev_60,Zubarev_74},
let us define the retarded function 
\begin{equation}
G(i,j)=\langle R[\psi(i)\psi^{\dagger}(j)]\rangle=\theta(t_{i}-t_{j})\langle\{\psi(i),\,\psi^{\dagger}(j)\}\rangle\label{3.13}
\end{equation}

By means of the Heisenberg equation (\ref{3.5}) and using the decomposition
(\ref{3.7}), the Green's function $G(i,j)$ satisfies the equation
\begin{equation}
\Lambda(\partial_{i})G(i,j)\Lambda^{\dagger}(\overleftarrow{\partial}_{j})=\Lambda(\partial_{i})G_{0}(i,j)\Lambda^{\dagger}(\overleftarrow{\partial}_{j})+\langle R[\delta J(i)\delta J^{\dagger}(j)]\rangle\label{3.14}
\end{equation}
where the derivative operator $\Lambda(\partial_{i})$ is defined
as 
\begin{equation}
\Lambda(\partial_{i})=\mathrm{i}\frac{\partial}{\partial t_{i}}-\varepsilon(-\mathrm{i}\nabla_{i})\label{3.15}
\end{equation}
and the propagator $G^{0}(i,j)$ is defined by the equation 
\begin{equation}
\Lambda(\partial_{i})G^{0}(i,j)=\mathrm{i}\delta(t_{i}-t_{j})I(i,j)\label{3.16}
\end{equation}

By introducing the Fourier transform 
\begin{equation}
G(i,j)=\frac{1}{N}\sum_{\mathbf{k}}\frac{\mathrm{i}}{2\pi}\int d\omega\mathrm{e}^{\mathrm{i}\mathbf{k}\cdot(\mathbf{R_{i}}-\mathbf{R_{j}})-\mathrm{i}\omega(t_{i}-t_{j})}G(\mathbf{k},\omega)\label{3.17}
\end{equation}
equation (\ref{3.14}) in momentum space can be written as 
\begin{equation}
G(\mathbf{k},\omega)=G^{0}(\mathbf{k},\omega)+G^{0}(\mathbf{k},\omega)I^{-1}(\mathbf{k})\Sigma(\mathbf{k},\omega)G(\mathbf{k},\omega)\label{3.18}
\end{equation}
and can be formally solved as 
\begin{equation}
G(\mathbf{k},\omega)=\frac{1}{\omega-\varepsilon(\mathbf{k})-\Sigma(\mathbf{k},\omega)}I(\mathbf{k})\label{3.19}
\end{equation}
where the self-energy $\Sigma(\mathbf{k},\omega)$ has the expression
\begin{equation}
\Sigma(\mathbf{k},\omega)=B_{irr}(\mathbf{k},\omega)I^{-1}(\mathbf{k})\label{3.20}
\end{equation}
with 
\begin{equation}
B(\mathbf{k},\omega)=\mathcal{F}\langle R[\delta J(i)\delta J^{\dagger}(j)]\rangle\label{3.21}
\end{equation}

The notation $\mathcal{F}$ denotes the Fourier transform and the
subscript $irr$ indicates that the irreducible part of the propagator
$B(\mathbf{k},\omega)$ is taken. Equation (\ref{3.18}) is nothing
else than the Dyson equation for composite fields and represents the
starting point for a perturbative calculation in terms of the propagator
$G^{0}(\mathbf{k},\omega)$. This quantity will be calculated in the
next section. Then, the attention will be given to the calculation
of the self-energy $\Sigma(\mathbf{k},\omega)$. It should be noted
that the computation of the two quantities $G^{0}(\mathbf{k},\omega)$
and $\Sigma(\mathbf{k},\omega)$ are intimately related. The total
weight of the self-energy corrections is bounded by the weight of
the residual source operator $\delta J(i)$. According to this, it
can be made smaller and smaller by increasing the components of the
basis $\psi(i)$ {[}e.g., by including higher-order composite operators
appearing in $\delta J(i)${]}. The result of such a procedure will
be the inclusion in the energy matrix of part of the self-energy as
an expansion in terms of coupling constants multiplied by the weights
of the newly included basis operators. In general, the enlargement
of the basis leads to a new self-energy with a smaller total weight.
However, it is necessary pointing out that this process can be quite
cumbersome and the inclusion of fully momentum and frequency dependent
self-energy corrections can be necessary to effectively take into
account low-energy and virtual processes. According to this, one can
choose a reasonable number of components for the basic set and then
use another approximation method to evaluate the residual dynamical
corrections.

\subsection{Two-pole Approximation\label{sec:Two-pole-Approximation}}

According to equation (\ref{3.16}), the free propagator $G^{0}(\mathbf{k},\omega)$
is determined by the following expression 
\begin{equation}
G^{0}(\mathbf{k},\omega)=\frac{1}{\omega-\varepsilon(\mathbf{k})}I(\mathbf{k})\label{3.22}
\end{equation}

For a paramagnetic state, straightforward calculations give the following
expressions for the normalization $I(\mathbf{k})$ and energy $\varepsilon(\mathbf{k})$
matrices 
\begin{equation}
I(\mathbf{k})=\left(\begin{array}{cc}
1-n/2 & 0\\
0 & n/2
\end{array}\right)=\left(\begin{array}{cc}
I_{11} & 0\\
0 & I_{22}
\end{array}\right)\label{3.23}
\end{equation}
\begin{equation}
\begin{array}{l}
\varepsilon_{11}(\mathbf{k})=-\mu-4tI_{11}^{-1}[\Delta+(1-n+p)\alpha(\mathbf{k})]\\
\varepsilon_{12}(\mathbf{k})=4tI_{22}^{-1}[\Delta+(p-I_{22})\alpha(\mathbf{k})]\\
\varepsilon_{21}(\mathbf{k})=4tI_{11}^{-1}[\Delta+(p-I_{22})\alpha(\mathbf{k})]\\
\varepsilon_{22}(\mathbf{k})=U-\mu-4tI_{22}^{-1}[\Delta+p\alpha(\mathbf{k})]
\end{array}\label{3.24}
\end{equation}
where $n=\langle n(i)\rangle$ is the filling and 
\begin{equation}
\begin{array}{l}
\Delta=\langle\xi^{\alpha}(i)\xi^{\dagger}(i)\rangle-\langle\eta^{\alpha}(i)\eta^{\dagger}(i)\rangle\\
p=\frac{1}{4}\langle n_{\mu}^{\alpha}(i)n_{\mu}(i)\rangle-\langle[c_{\uparrow}(i)c_{\downarrow}(i)]^{\alpha}c_{\downarrow}^{\dagger}(i)c_{\uparrow}^{\dagger}(i)\rangle
\end{array}\label{3.25}
\end{equation}
Then, (\ref{3.22}) can be written in spectral form as 
\begin{equation}
G^{0}(\mathbf{k},\omega)=\sum_{n=1}^{2}\frac{\sigma^{(n)}(\mathbf{k})}{\omega-E_{n}(\mathbf{k})+\mathrm{i}\delta}\label{3.26}
\end{equation}

The energy spectra $E_{n}(\mathbf{k})$ and the spectral functions
$\sigma^{(n)}(\mathbf{k})$ are given by 
\begin{equation}
E_{1}(\mathbf{k})=R(\mathbf{k})+Q(\mathbf{k})\quad\quad\quad\quad E_{2}(\mathbf{k})=R(\mathbf{k})-Q(\mathbf{k})\label{3.27}
\end{equation}
\begin{equation}
\begin{array}{l}
\sigma_{11}^{(1)}(\mathbf{k})=\frac{I_{11}}{2}\left[1+\frac{g(\mathbf{k})}{2Q(\mathbf{k})}\right]\\
\sigma_{12}^{(1)}(\mathbf{k})=\frac{m_{12}(\mathbf{k})}{2Q(\mathbf{k})}\\
\sigma_{22}^{(1)}(\mathbf{k})=\frac{I_{22}}{2}\left[1-\frac{g(\mathbf{k})}{2Q(\mathbf{k})}\right]
\end{array}\quad\quad\quad\quad\quad\begin{array}{l}
\sigma_{11}^{(2)}(\mathbf{k})=\frac{I_{11}}{2}\left[1-\frac{g(\mathbf{k})}{2Q(\mathbf{k})}\right]\\
\sigma_{12}^{(2)}(\mathbf{k})=-\frac{m_{12}(\mathbf{k})}{2Q(\mathbf{k})}\\
\sigma_{22}^{(2)}(\mathbf{k})=\frac{I_{22}}{2}\left[1+\frac{g(\mathbf{k})}{2Q(\mathbf{k})}\right]
\end{array}\label{3.28}
\end{equation}
where 
\begin{equation}
\begin{array}{l}
R(\mathbf{k})=-\mu-4t\alpha(\mathbf{k})+\frac{1}{2}U-\frac{\varepsilon_{12}(\mathbf{k})}{2I_{11}}\\
Q(\mathbf{k})=\frac{1}{2}\sqrt{g^{2}(\mathbf{k})+\frac{4\varepsilon_{12}^{2}(\mathbf{k})I_{22}}{I_{11}}}\\
g(\mathbf{k})=-U+\frac{1-n}{I_{11}}\varepsilon_{12}(\mathbf{k})
\end{array}\label{3.29}
\end{equation}

The energy matrix $\varepsilon(\mathbf{k})$ contains three parameters:
$\mu$, the chemical potential, $\Delta$, the difference between
upper and lower intra-subband contributions to kinetic energy, and
$p$, a combination of the nearest-neighbor charge-charge, spin-spin
and pair-pair correlation functions. These parameters will be determined
in a self-consistent way by means of algebra constraints in terms
of the external parameters $n$, $U$, and $T$.

\subsection{Non-Crossing Approximation\label{s3.3}}

The calculation of the self-energy $\Sigma(\mathbf{k},\omega)$ requires
the calculation of the higher-order propagator $B(\mathbf{k},\omega)$
{[}cfr. (\ref{3.21}){]}. We shall compute this quantity by using
the Non-Crossing Approximation (NCA). By neglecting the pair term
$c(i)c^{\dagger\alpha}(i)c(i)$, the source $J(i)$ can be written
as 
\begin{equation}
J(\mathbf{i},t)=\sum_{\mathbf{j}}a(\mathbf{i,j},t)\psi(\mathbf{j},t)\label{3.30}
\end{equation}
where 
\begin{equation}
\begin{array}{l}
a_{11}(\mathbf{i,j},t)=-\mu\delta_{\mathbf{ij}}-4t\alpha_{\mathbf{ij}}-2t\sigma^{\mu}n_{\mu}(i)\alpha_{\mathbf{ij}}\\
a_{12}(\mathbf{i,j},t)=-4t\alpha_{\mathbf{ij}}-2t\sigma^{\mu}n_{\mu}(i)\alpha_{\mathbf{ij}}\\
a_{21}(\mathbf{i,j},t)=2t\sigma^{\mu}n_{\mu}(i)\alpha_{\mathbf{ij}}\\
a_{22}(\mathbf{i,j},t)=(U-\mu)\delta_{\mathbf{ij}}+2t\sigma^{\mu}n_{\mu}(i)\alpha_{\mathbf{ij}}
\end{array}\label{3.31}
\end{equation}

Then, for the calculation of $B_{irr}(i,j)=\langle R[\delta J(i)\delta J^{\dagger}(j)]\rangle_{irr}$,
we approximate 
\begin{equation}
\delta J(\mathbf{i},t)\approx\sum_{\mathbf{j}}\left[a(\mathbf{i,j},t)-\langle a(\mathbf{i,j},t)\rangle\right]\psi(\mathbf{j},t)\label{3.32}
\end{equation}

Therefore 
\begin{equation}
B_{irr}(i,j)=4t^{2}F(i,j)(1-\sigma_{1})\label{3.33}
\end{equation}
where we defined 
\begin{equation}
F(i,j)=\langle R[\sigma^{\mu}\delta n_{\mu}(i)c^{\alpha}(i)c^{\dagger\alpha}(j)\delta n_{\lambda}(j)\sigma^{\lambda}]\rangle\label{3.34}
\end{equation}
with $\delta n_{\mu}(i)=n_{\mu}(i)-\langle n_{\mu}(i)\rangle$. The
self-energy (\ref{3.20}) is written as 
\begin{equation}
\Sigma(\mathbf{k},\omega)=4t^{2}F(\mathbf{k},\omega)\left(\begin{array}{cc}
I_{11}^{-2} & -I_{11}^{-1}I_{22}^{-1}\\
-I_{11}^{-1}I_{22}^{-1} & I_{22}^{-2}
\end{array}\right)\label{3.35}
\end{equation}

In order to calculate the retarded function $F(i,j)$, first we use
the spectral theorem to express 
\begin{equation}
F(i,j)=\frac{i}{2\pi}\int_{-\infty}^{+\infty}d\omega\mathrm{e}^{-\mathrm{i}\omega(t_{i}-t_{j})}\frac{1}{2\pi}\int_{-\infty}^{+\infty}d\omega'\frac{1+\mathrm{e}^{-\beta\omega'}}{\omega-\omega'+\mathrm{i}\varepsilon}C(\mathbf{i-j},\omega')\label{3.36}
\end{equation}
where $C(\mathbf{i-j},\omega')$ is the correlation function 
\begin{equation}
C(i,j)=\langle\sigma^{\mu}\delta n_{\mu}(i)c^{\alpha}(i)c^{\dagger\alpha}(j)\delta n_{\lambda}(j)\sigma^{\lambda}\rangle=\frac{1}{2\pi}\int d\omega\mathrm{e}^{-\mathrm{i}\omega(t_{i}-t_{j})}C(\mathbf{i-j},\omega)\label{3.37}
\end{equation}

Next, we use the Non-Crossing Approximation (NCA) and approximate
\begin{equation}
\langle\sigma^{\mu}\delta n_{\mu}(i)c^{\alpha}(i)c^{\dagger\alpha}(j)\delta n_{\lambda}(j)\sigma^{\lambda}\rangle\approx\langle\delta n_{\mu}(i)\delta n_{\mu}(j)\rangle\langle c^{\alpha}(i)c^{\dagger\alpha}(j)\rangle\label{3.38}
\end{equation}

By means of this decoupling and using again the spectral theorem we
finally have 
\begin{equation}
\begin{array}{l}
F(\mathbf{k},\omega)=\frac{1}{\pi}\int_{-\infty}^{+\infty}d\omega'\frac{1}{\omega-\omega'+\mathrm{i}\delta}\frac{a^{2}}{(2\pi)^{3}}\int d^{2}pd\Omega\alpha^{2}(p)\\
\quad\quad\quad\quad\times\left[\tanh\frac{\beta\Omega}{2}+\coth\frac{\beta(\omega'-\Omega)}{2}\right]\Im[G_{cc}(\mathbf{p},\Omega)]\Im[\chi(\mathbf{k-p},\omega'-\Omega)]
\end{array}\label{3.39}
\end{equation}
where $G_{cc}(\mathbf{k},\omega)$ is the retarded electronic Green's
function {[}cfr. (\ref{3.13}){]} 
\begin{equation}
G_{cc}(\mathbf{k},\omega)=\sum_{a,b=1}^{2}G_{ab}(\mathbf{k},\omega)\label{3.40}
\end{equation}
and 
\begin{equation}
\chi(\mathbf{k},\omega)=\sum_{\mu}\mathcal{F}\left\langle R\left[\delta n_{\mu}(i)\delta n_{\mu}(j)\right]\right\rangle \label{3.41}
\end{equation}
is the total charge and spin dynamical susceptibility. The result
(\ref{3.39}) shows that the calculation of the self-energy requires
the knowledge of the bosonic propagator (\ref{3.41}). This problem
will be considered in the following section.

It is worth noting that the NCA can also be applied to the casual
propagators giving the same result. In general, the knowledge of the
self-energy requires the calculation of the higher-order propagator
$B^{Q}(i,j)=\langle{\cal Q}[\delta J(i)\delta J^{\dagger}(j)]\rangle$,
where $Q$ can be $R$ (retarded propagator) o $T$ (causal propagator).
Then, typically we have to calculate propagator of the form

\begin{equation}
H^{R}(i,j)=\langle{\cal R}[B(i)F(i)F^{\dagger}(j)B^{\dagger}(j)]\rangle
\end{equation}
where $F(i)$ and $B(i)$ are fermionic and bosonic field operators,
respectively. By means of the spectral representation we can write

\begin{equation}
H^{R}({\bf k},\omega)=-\frac{1}{\pi}\int_{-\infty}^{+\infty}d\omega'\frac{1}{\omega-\omega'+{\rm i}\delta}\coth\frac{\beta\omega'}{2}\Im[H^{C}({\bf k},\omega')]
\end{equation}
where $H^{C}(i,j)=\langle{\cal T}[B(i)F(i)F^{\dagger}(j)B^{\dagger}(j)]\rangle$
is the causal propagator. In the NCA, we approximate 
\begin{equation}
H^{C}(i,j)\approx f^{C}(i,j)b^{C}(i,j)\quad\quad\begin{array}{c}
f^{C}(i,j)=\langle{\cal T}[F(i)F^{\dagger}(j)]\rangle\\
\\
b^{C}(i,j)=\langle{\cal T}[B(i)B^{\dagger}(j)]\rangle
\end{array}
\end{equation}
Then, we can use the spectral representation to obtain
\begin{eqnarray}
f^{C}({\bf k},\omega) & = & -\frac{1}{\pi}\int_{-\infty}^{+\infty}d\omega'[\frac{1-f_{{\rm F}}(\omega')}{\omega-\omega'+{\rm i}\delta}+\frac{f_{{\rm F}}(\omega')}{\omega-\omega'-{\rm i}\delta}]\Im[f^{R}({\bf k},\omega')]\\
b^{C}({\bf k},\omega) & = & -\frac{1}{\pi}\int_{-\infty}^{+\infty}d\omega'[\frac{1+f_{B}(\omega')}{\omega-\omega'+{\rm i}\delta}-\frac{f_{B}(\omega')}{\omega-\omega'-{\rm i}\delta}]\Im[b^{R}({\bf k},\omega')]
\end{eqnarray}
 which leads to
\begin{eqnarray}
H^{R}({\bf k},\omega) & = & \frac{1}{\pi}\int_{-\infty}^{+\infty}d\omega'\frac{1}{\omega-\omega'+{\rm i}\delta}\frac{a^{d}}{(2\pi)^{d+1}}\int_{\Omega_{B}}d^{d}pd\Omega\Im[f^{R}(p,\Omega)]\nonumber \\
 & \times & \Im[b^{R}(k-p,\omega'-\Omega)][\tanh\frac{\beta\Omega}{2}+\coth\frac{\beta(\omega'-\Omega)}{2}]
\end{eqnarray}

It is worth noting that, up to this point, the system of equations
for the Green's function and the \emph{anomalous }self-energy is similar
to the one derived in the two-particle self-consistent approach (TPSC)
\cite{Vilk_95,Tremblay_06}, the DMFT$+\Sigma$ approach \cite{Sadovskii_01,Sadovskii_05,Kuchinskii_05,Kuchinskii_06,Kuchinskii_06a}
and a Mori-like approach by Plakida and coworkers \cite{Plakida_01,Plakida_06,Plakida_10}.
It would be the way to compute the dynamical spin and charge susceptibilities
to be completely different as, instead of relying on a phenomenological
model and neglecting the charge susceptibility as these approches
do, we will use a self-consistent two-pole approximation. Obviously,
a proper description of the spin and charge dynamics would definitely
require the inclusion of a proper self-energy term in the charge and
spin propagators too in order to go beyond both any phenomenological
approch and the two-pole approximation (in preparation). On the other
hand, the description of the electronic anomalous features could (actually
will, see in the following) not need this further, and definitely
not trivial, complication.

\subsection{Dynamical susceptibility\label{s3.4}}

In this section, we shall present a calculation of the charge-charge
and spin-spin propagators (\ref{3.41}) within the two-pole approximation.
This approximation has shown to be capable to catch correctly some
of the physical features of Hubbard model dynamics (for all details
see Ref.~\cite{Avella_03}).

Let us define the composite bosonic field 
\begin{equation}
N^{(\mu)}(i)=\left(\begin{array}{c}
n_{\mu}(i)\\
\rho_{\mu}(i)
\end{array}\right)\quad\quad\quad\begin{array}{l}
n_{\mu}(i)=c^{\dagger}(i)\sigma_{\mu}c(i)\\
\rho_{\mu}(i)=c^{\dagger}(i)\sigma_{\mu}c^{\alpha}(i)-c^{\dagger\alpha}(i)\sigma_{\mu}c(i)
\end{array}\label{3.42}
\end{equation}

This field satisfies the Heisenberg equation 
\begin{equation}
\mathrm{i}\frac{\partial}{\partial t}N^{(\mu)}(i)=J^{(\mu)}(i)=\left(\begin{array}{c}
J_{1}^{(\mu)}(i)\\
J_{2}^{(\mu)}(i)
\end{array}\right)\quad\quad\quad\begin{array}{l}
J_{1}^{(\mu)}(i)=-4t\rho_{\mu}(i)\\
J_{2}^{(\mu)}(i)=U\kappa_{\mu}(i)-4tl_{\mu}(i)
\end{array}\label{3.43}
\end{equation}
where the higher-order composite fields $\kappa_{\mu}(i)$ and $l_{\mu}(i)$
are defined as 
\begin{equation}
\begin{array}{l}
\kappa_{\mu}(i)=c^{\dagger}(i)\sigma_{\mu}\eta^{\alpha}(i)-\eta^{\dagger}(i)\sigma_{\mu}c^{\alpha}(i)+\eta^{\dagger\alpha}(i)\sigma_{\mu}c(i)-c^{\dagger\alpha}(i)\sigma_{\mu}\eta(i)\\
l_{\mu}(i)=c^{\dagger}(i)\sigma_{\mu}c^{\alpha^{2}}(i)+c^{\dagger\alpha^{2}}(i)\sigma_{\mu}c(i)-2c^{\dagger\alpha}(i)\sigma_{\mu}c^{\alpha}(i)
\end{array}\label{3.44}
\end{equation}
and we are using the notation 
\begin{equation}
c^{\alpha^{2}}(\mathbf{i},t)=\sum_{\mathbf{j}}\alpha_{\mathbf{ij}}^{2}c(\mathbf{j},t)=\sum_{\mathbf{jl}}\alpha_{\mathbf{il}}\alpha_{\mathbf{lj}}c(\mathbf{j},t)\label{3.45}
\end{equation}

We linearize the equation of motion (\ref{3.43}) for the composite
field $N^{(\mu)}(i)$ by using the same criterion as in Section \ref{sec2.1}
(i.e., the neglected residual current $\delta J^{(\mu)}(i)$ is orthogonal
to the chosen basis (\ref{3.42})) 
\begin{equation}
\mathrm{i}\frac{\partial}{\partial t}N^{(\mu)}(\mathbf{i},t)=\sum_{\mathbf{j}}\varepsilon^{(\mu)}(\mathbf{i,j})N^{(\mu)}(\mathbf{j},t)\label{3.46}
\end{equation}
where the energy matrix is given by 
\begin{equation}
m^{(\mu)}(\mathbf{i,j})=\sum_{\mathbf{l}}\varepsilon^{(\mu)}(\mathbf{i,l})I^{(\mu)}(\mathbf{l,j})\label{3.47}
\end{equation}
and the normalization matrix $I^{(\mu)}$ and the $m^{(\mu)}$-matrix
have the following definitions 
\begin{equation}
I^{(\mu)}(\mathbf{i,j})=\langle[N^{(\mu)}(\mathbf{i},t),\, N^{(\mu)\dagger}(\mathbf{j},t)]\rangle\label{3.48}
\end{equation}
\begin{equation}
m^{(\mu)}(\mathbf{i,j})=\langle[J^{(\mu)}(\mathbf{i},t),\, N^{(\mu)\dagger}(\mathbf{j},t)]\rangle\label{3.49}
\end{equation}
As it can be easily verified, in the paramagnetic phase the normalization
matrix $I^{(\mu)}$ does not depend on the index $\mu$: charge and
spin operators have the same weight. The two matrices $I^{(\mu)}$
and $m^{(\mu)}$ have the following form in momentum space 
\begin{equation}
I^{(\mu)}(\mathbf{k})=\left(\begin{array}{cc}
0 & I_{12}^{(\mu)}(\mathbf{k})\\
I_{12}^{(\mu)}(\mathbf{k}) & 0
\end{array}\right)\label{3.50}
\end{equation}
\begin{equation}
m^{(\mu)}(\mathbf{k})=\left(\begin{array}{cc}
m_{11}^{(\mu)}(\mathbf{k}) & 0\\
0 & m_{22}^{(\mu)}(\mathbf{k})
\end{array}\right)\label{3.51}
\end{equation}
where 
\begin{equation}
\begin{array}{l}
I_{12}^{(\mu)}(\mathbf{k})=4[1-\alpha(\mathbf{k})]C_{cc}^{\alpha}\\
m_{11}^{(\mu)}(\mathbf{k})=-4tI_{12}^{(\mu)}(\mathbf{k})\\
m_{22}^{(\mu)}(\mathbf{k})=-4tI_{l_{\mu}\rho_{\mu}}(\mathbf{k})+UI_{\kappa_{\mu}\rho_{\mu}}(\mathbf{k})
\end{array}\label{3.52}
\end{equation}

The parameter $C^{\alpha}$ is the electronic correlation function
$C^{\alpha}=\langle c^{\alpha}(i)c^{\dagger}(i)\rangle$. The quantities
$I_{l_{\mu}\rho_{\mu}}(\mathbf{k})$ and $I_{\kappa_{\mu}\rho_{\mu}}(\mathbf{k})$
are defined as 
\begin{equation}
I_{l_{\mu}\rho_{\mu}}(\mathbf{k})=\mathcal{F}\langle[l_{\mu}(\mathbf{i},t),\,\rho_{\mu}^{\dagger}(\mathbf{j},t)]\rangle\quad\quad\quad I_{\kappa_{\mu}\rho_{\mu}}(\mathbf{k})=\mathcal{F}\langle[\kappa_{\mu}(\mathbf{i},t),\,\rho_{\mu}^{\dagger}(\mathbf{j},t)]\rangle\label{3.53}
\end{equation}
Let us define the causal Green's function (for bosonic-like fields
we have to compute the casual Green's function and deduce from this
latter the retarded one according to the prescriptions in Ref.~\cite{Theory,Avella_11a})
\begin{eqnarray}
G^{(\mu)}(i,j) & = & \langle T[N^{(\mu)}(i)N^{(\mu)\dagger}(j)]\rangle\nonumber \\
 & = & \frac{\mathrm{i}a^{2}}{(2\pi)^{3}}\int d^{2}k\, d\omega\,\mathrm{e}^{\mathrm{i}\mathbf{k}\cdot(\mathbf{R_{i}-R_{j}})-\mathrm{i}\omega(t_{i}-t_{j})}G^{(\mu)}(\mathbf{k,\omega})\label{3.54}
\end{eqnarray}
By means of the equation of motion (\ref{3.46}), the Fourier transform
of $G^{(\mu)}(i,j)$ satisfies the following equation 
\begin{equation}
[\omega-\varepsilon^{(\mu)}(\mathbf{k})]G^{(\mu)}(\mathbf{k},\omega)=I^{(\mu)}(\mathbf{k})\label{3.55}
\end{equation}
where the energy matrix has the explicit form 
\begin{equation}
\varepsilon^{(\mu)}(\mathbf{k})=\left(\begin{array}{cc}
0 & \varepsilon_{12}^{(\mu)}(\mathbf{k})\\
\varepsilon_{21}^{(\mu)}(\mathbf{k}) & 0
\end{array}\right)\quad\quad\begin{array}{l}
\varepsilon_{12}^{(\mu)}(\mathbf{k})=-4t\\
\varepsilon_{21}^{(\mu)}(\mathbf{k})=m_{22}^{(\mu)}(\mathbf{k})/I_{12}^{(\mu)}(\mathbf{k})
\end{array}\label{3.56}
\end{equation}
The solution of (\ref{3.55}) is 
\begin{eqnarray}
G^{(\mu)}(\mathbf{k},\omega) & = & \Gamma^{(\mu)}(\mathbf{k})\left[\frac{1}{\omega+\mathrm{i}\delta}-\frac{1}{\omega-\mathrm{i}\delta}\right]\nonumber \\
 & + & \sum_{n=1}^{2}\sigma^{(n,\mu)}(\mathbf{k})\left[\frac{1+f_{\mathrm{B}}(\omega)}{\omega-\omega_{n}^{(\mu)}(\mathbf{k})+\mathrm{i}\delta}-\frac{f_{\mathrm{B}}(\omega)}{\omega-\omega_{n}^{(\mu)}(\mathbf{k})-\mathrm{i}\delta}\right]\label{3.57}
\end{eqnarray}
where $\Gamma^{(\mu)}(\mathbf{k})$ is the zero frequency function
($2\times2$ matrix) \cite{Theory,Avella_11a} and $f_{\mathrm{B}}(\omega)=[\mathrm{e}^{\beta\omega}-1]^{-1}$
is the Bose distribution function. Correspondingly, the correlation
function $C^{(\mu)}(\mathbf{k},\omega)=\langle N^{(\mu)}(i)N^{(\mu)\dagger}(j)\rangle$
has the expression 
\begin{equation}
C^{(\mu)}(\mathbf{k},\omega)=2\pi\Gamma^{(\mu)}(\mathbf{k})\delta(\omega)+2\pi\sum_{n=1}^{2}\delta[\omega-\omega_{n}^{(\mu)}(\mathbf{k})][1+f_{\mathrm{B}}(\omega)]\sigma^{(n,\mu)}(\mathbf{k})\label{3.58}
\end{equation}
The energy spectra $\omega_{n}^{(\mu)}(\mathbf{k})$ are given by
\begin{equation}
\begin{array}{l}
\omega_{n}^{(\mu)}(\mathbf{k})=(-)^{n}\omega^{(\mu)}(\mathbf{k})\\
\omega^{(\mu)}(\mathbf{k})=\sqrt{\varepsilon_{12}^{(\mu)}(\mathbf{k})\varepsilon_{21}^{(\mu)}(\mathbf{k})}
\end{array}\label{3.59}
\end{equation}
and the spectral functions $\sigma^{(n,\mu)}(\mathbf{k})$ have the
following expression 
\begin{equation}
\sigma^{(n,\mu)}(\mathbf{k})=\frac{{I_{12}^{(\mu)}(\mathbf{k})}}{2}\left(\begin{array}{cc}
\frac{\varepsilon_{12}^{(\mu)}(\mathbf{k})}{\omega_{n}^{(\mu)}(\mathbf{k})} & 1\\
1 & \frac{\varepsilon_{21}^{(\mu)}(\mathbf{k})}{\omega_{n}^{(\mu)}(\mathbf{k})}
\end{array}\right)\label{3.60}
\end{equation}
Straightforward but lengthy calculations (see Ref.~\cite{Avella_03})
give for the 2D system the following expressions for the commutators
in (\ref{3.53}) 
\begin{equation}
\begin{array}{l}
I_{l_{\mu}\rho_{\mu}}(\mathbf{k})=\frac{3}{4}[1-\alpha(\mathbf{k})](12C^{\alpha}+C^{\lambda}+6C^{\mu})-3[1-\beta(\mathbf{k})](C^{\alpha}+C^{\mu})\\
-\frac{3}{4}[1-\eta(\mathbf{k})](C^{\alpha}+C^{\lambda}+2C^{\mu})+\frac{1}{4}[1-\lambda(\mathbf{k})]C^{\lambda}+\frac{3}{2}[1-\mu(\mathbf{k})]C^{\mu}
\end{array}\label{3.61}
\end{equation}
\begin{equation}
\begin{array}{l}
I_{\kappa_{\mu}\rho_{\mu}}(\mathbf{k})=-2[1-\alpha(\mathbf{k})]D+[1-2\alpha(\mathbf{k})](2E^{\beta}+E^{\eta})+2\beta(\mathbf{k})E^{\beta}+\eta(\mathbf{k})E^{\eta}\\
+[1-2\alpha(\mathbf{k})]a_{\mu}+\frac{1}{4}[b_{\mu}+2\beta(\mathbf{k})c_{\mu}+\eta(\mathbf{k})d_{\mu}]
\end{array}\label{3.62}
\end{equation}
where $\alpha(\mathbf{k})$, $\beta(\mathbf{k})$, $\eta(\mathbf{k})$,
$\mu(\mathbf{k})$, and $\lambda(\mathbf{k})$ are the Fourier transforms
of the projectors on the first, second, third, fourth, and fifth nearest-neighbor
sites. The parameters appearing in (\ref{3.61}) and (\ref{3.62})
are defined by 
\begin{equation}
\begin{array}{ll}
E=\langle c(i)\eta^{\dagger}(i)\rangle & C^{\alpha}=\langle c^{\alpha}(i)c^{\dagger}(i)\rangle\\
E^{\beta}=\langle c^{\beta}(i)\eta^{\dagger}(i)\rangle & C^{\lambda}=\langle c^{\lambda}(i)c^{\dagger}(i)\rangle\\
E^{\eta}=\langle c^{\beta}(i)\eta^{\dagger}(i)\rangle & C^{\mu}=\langle c^{\mu}(i)c^{\dagger}(i)\rangle
\end{array}\label{3.63}
\end{equation}
\begin{equation}
\begin{array}{l}
a_{\mu}=2\langle c^{\dagger}(i)\sigma_{\mu}c^{\alpha}(i)c^{\dagger}(i)\sigma_{\mu}c^{\alpha}(i)\rangle-\langle c^{\dagger\alpha}(i)\sigma_{\mu}\sigma^{\lambda}\sigma_{\mu}c^{\alpha}(i)n_{\lambda}(i)\rangle\\
b_{\mu}=2\langle c^{\dagger}(i)\sigma_{\mu}c^{\dagger}(i)\sigma_{\mu}[c(i)c(i)]^{\alpha}\rangle-\langle c^{\dagger}(i)\sigma_{\mu}\sigma^{\lambda}\sigma_{\mu}c(i)n_{\lambda}^{\alpha}(i)\rangle\\
c_{\mu}=2\langle c^{\dagger}(i)\sigma_{\mu}c^{\dagger}(i^{\eta})\sigma_{\mu}c(i^{\alpha})c(i^{\alpha})\rangle-\langle c^{\dagger}(i)\sigma_{\mu}\sigma^{\lambda}\sigma_{\mu}c(i^{\eta})n_{\lambda}(i^{\alpha})\rangle\\
d_{\mu}=2\langle c^{\dagger}(i)\sigma_{\mu}c^{\dagger}(i^{\beta})\sigma_{\mu}c(i^{\alpha})c(i^{\alpha})\rangle-\langle c^{\dagger}(i)\sigma_{\mu}\sigma^{\lambda}\sigma_{\mu}c(i^{\beta})n_{\lambda}(i^{\alpha})\rangle
\end{array}\label{3.64}
\end{equation}
where we used the notation 
\begin{equation}
\begin{array}{l}
i=(i_{x},i_{y},t)\\
i^{\alpha}=(i_{x}+a,i_{y},t)
\end{array}\quad\quad\quad\begin{array}{l}
i^{\beta}=(i_{x}+a,i_{y}+a,t)\\
i^{\eta}=(i_{x}+2a,i_{y},t)
\end{array}\label{3.65}
\end{equation}

We see that the bosonic Green's function $G^{(\mu)}(i,j)=\langle T[N^{(\mu)}(i)N^{(\mu)\dagger}(j)]\rangle$
depends on the following set of parameters. Fermionic correlators:
$C^{\alpha}$, $C^{\lambda}$, $C^{\mu}$, $E^{\beta}$, $E^{\eta}$,
$D$; bosonic correlators: $a_{\mu}$, $b_{\mu}$, $c_{\mu}$, $d_{\mu}$;
zero frequency matrix $\Gamma^{(\mu)}(\mathbf{k})$. The fermionic
parameters are calculated through the Fermionic correlation function
$C(i,j)=\langle\psi(i)\psi^{\dagger}(j)\rangle$. The bosonic parameters
are determined through symmetry requirements. In particular, the requirement
that the continuity equation be satisfied and that the susceptibility
be a single-value function at $\mathbf{k=0}$ leads to the following
equations 
\begin{equation}
\begin{array}{l}
b_{\mu}=a_{\mu}+3D+2E^{\beta}+E^{\eta}-6\frac{t}{U}\left(C^{\alpha}+C^{\lambda}-2C^{\mu}\right)\\
c_{\mu}=a_{\mu}-D-2E^{\beta}+E^{\eta}+6\frac{t}{U}\left(C^{\alpha}+C^{\lambda}-2C^{\mu}\right)\\
d_{\mu}=a_{\mu}-D+2E^{\beta}-3E^{\eta}-6\frac{t}{U}\left(C^{\alpha}+C^{\lambda}-2C^{\mu}\right)
\end{array}\label{3.66}
\end{equation}

The remaining parameters $a_{\mu}$ and $\Gamma_{11}^{(\mu)}(\mathbf{k})$
are fixed by means of the Pauli principle 
\begin{equation}
\langle n_{\mu}(i)n_{\mu}(i)\rangle=\left\{ \begin{array}{l}
n+2D\quad\quad\textrm{for}\quad\mu=0\\
n-2D\quad\quad\textrm{for}\quad\mu=1,2,3
\end{array}\right.\label{3.67}
\end{equation}
where $D=\langle n_{\uparrow}(i)n_{\downarrow}(i)\rangle$ is the
double occupancy, and by the ergodic value 
\begin{equation}
\Gamma_{11}^{(\mu)}(\mathbf{k})=\delta_{\mu,0}\frac{{(2\pi)^{2}}}{{a^{2}}}\delta^{(2)}(\mathbf{k})\langle n\rangle^{2}\label{3.68}
\end{equation}
By putting (\ref{3.68}) into (\ref{3.57}) and (\ref{3.58}) we obtain
\begin{eqnarray}
\langle\delta n_{\mu}(i)\delta n_{\mu}(j)\rangle & = & \frac{a^{2}}{2(2\pi)^{2}}\sum_{n=1}^{2}\int d^{2}k\mathrm{e}^{\mathrm{i}\mathbf{k}\cdot(\mathbf{R_{i}-R_{j}})-\mathrm{i}\omega_{n}^{(\mu)}(\mathbf{k})(t_{i}-t_{j})}\label{3.69}\\
 & \times & \left[1+\coth\frac{\omega_{n}^{(\mu)}(\mathbf{k})}{2k_{\mathrm{B}}T}\right]\sigma_{11}^{(n,\mu)}(\mathbf{k})
\end{eqnarray}
\begin{eqnarray}
\langle R[\delta n_{\mu}(i)\delta n_{\mu}(j)]\rangle & = & \frac{\mathrm{i}a^{2}}{(2\pi)^{3}}\sum_{n=1}^{2}\int d^{2}kd\omega\mathrm{e}^{\mathrm{i}\mathbf{k}\cdot(\mathbf{R_{i}-R_{j}})-\mathrm{i}\omega(t_{i}-t_{j})}\label{3.70}\\
 & \times & \frac{\sigma_{11}^{(n,\mu)}(\mathbf{k})}{\omega-\omega_{n}^{(\mu)}(\mathbf{k})+\mathrm{i}\delta}
\end{eqnarray}

In conclusion, the dynamical susceptibility $\chi_{\mu}\left(\mathbf{k},\omega)\right)$,
which is independent from $\Gamma^{(\mu)}$ by construction, reads
as 
\begin{equation}
\chi_{\mu}\left(\mathbf{k},\omega\right)=-\mathcal{F}\left[\left\langle R\left[\delta n_{\mu}(i)\delta n_{\mu}(j)\right]\right\rangle \right]=\frac{16t[1-\alpha(\mathbf{k})]C^{\alpha}}{\omega^{2}-\left(\omega^{(\mu)}(\mathbf{k})\right)^{2}}\label{3.70bis}
\end{equation}
where no summation is implied on the $\mu$ index. It is really worth
noticing the very good agreement between the results we obtained within
this framework for the charge and spin dynamics of the Hubbard model
and the related numerical ones present in the literature (see Ref.~\cite{Avella_03}).

\begin{figure}
\noindent \centering{}\includegraphics[width=0.75\textwidth]{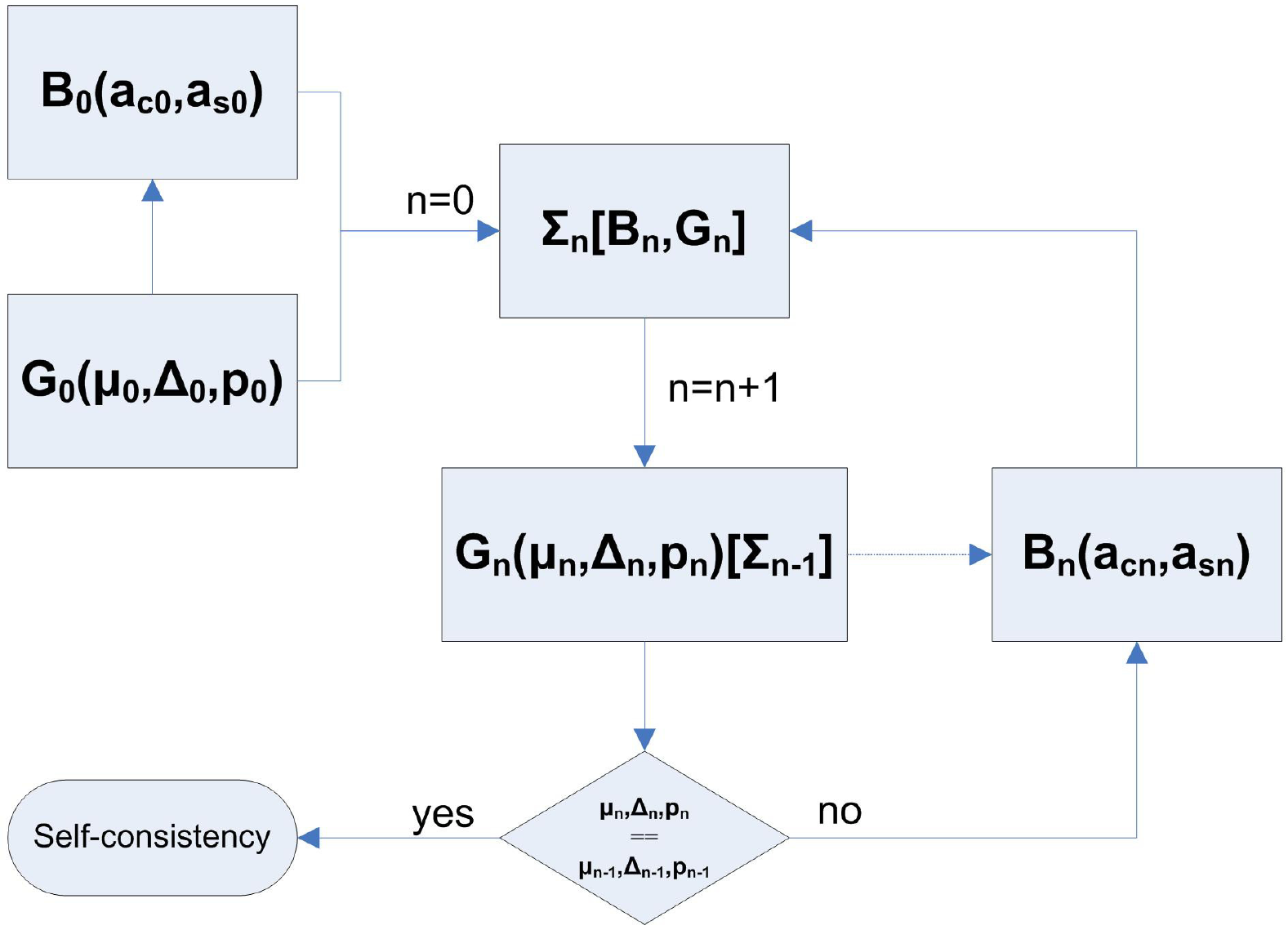}
\protect\caption{(Color online) Self-consistency scheme to compute the propagator $G$
in terms of the charge-charge and spin-spin propagator $B$ and the
residual self-energy $\Sigma$.\label{fig0}}
\end{figure}

\subsection{Self-consistency\label{s3.5}}

In this section, we will give a sketch of the procedure used to calculate
the Green's function $G(\mathbf{k},\omega)$. The starting point is
equation (\ref{3.19}), where the two matrices $I(\mathbf{k})$ and
$\varepsilon(\mathbf{k})$ are computed by means of the expressions
(\ref{3.23}) and (\ref{3.24}). The energy matrix $\varepsilon(\mathbf{k})$
depends on three parameters: $\mu$, $\Delta$, and $p$. To determine
these parameters we use the following set of algebra constraints 
\begin{equation}
\begin{array}{l}
n=2(1-C_{11}-C_{22})\\
\Delta=C_{11}^{\alpha}-C_{22}^{\alpha}\\
C_{12}=\langle\xi(i)\eta^{\dagger}(i)\rangle=0
\end{array}\label{3.71}
\end{equation}
where $C_{nm}$ and $C_{nm}^{\alpha}$ are the time-independent correlation
functions $C_{nm}=\langle\psi_{n}(i)\psi_{m}^{\dagger}(i)\rangle$
and $C_{nm}^{\alpha}=\langle\psi_{n}^{\alpha}(i)\psi_{m}^{\dagger}(i)\rangle$.
To calculate $\Sigma(\mathbf{k},\omega)$ we use the NCA; the results
given in Sec.~\ref{s3.3} show that within this approximation $\Sigma(\mathbf{k},\omega)$
is expressed in terms of the fermionic $G_{cc}(\mathbf{k},\omega)$
{[}cfr.~(\ref{3.40}){]} and of the bosonic $\chi_{\mu}(\mathbf{k},\omega)$
{[}cfr.~(\ref{3.41}){]} propagators. The bosonic propagator is calculated
within the two-pole approximation, using the expression (\ref{3.70}).
As shown in Sec.~\ref{s3.4}, $\chi_{\mu}(\mathbf{k},\omega)$ depends
on both electronic correlation functions {[}see (\ref{3.63}){]},
which can be straightforwardly computed from $G(\mathbf{k},\omega)$,
and bosonic correlation functions, one per each channel (charge and
spin), $a_{0}$ and $a_{3}$. The latter are determined by means of
the local algebra constraints (\ref{3.67}), where $n$ is the filling
and $D$ is the double occupancy, determined in terms of the electronic
correlation function as $D=n/2-C_{22}$.

According to this, the electronic Green's function $G(\mathbf{k},\omega)$
is computed through the self-consistency scheme depicted in Fig.~\ref{fig0}:
we first compute $G^{0}(\mathbf{k},\omega)$ and $\chi_{\mu}(\mathbf{k},\omega)$
in the two-pole approximation, then $\Sigma(\mathbf{k},\omega)$ and
consequently $G(\mathbf{k},\omega)$. Finally, we check how much the
fermionic parameters ($\mu$, $\Delta$, and $p$) changed and decide
if to stop or to continue by computing new $\chi_{_{\mu}}(\mathbf{k},\omega)$
and $\Sigma(\mathbf{k},\omega)$ after $G(\mathbf{k},\omega)$. Usually,
to get $6$ digits precision for fermionic parameters, we need $8$
full cycles to reach self-consistency on a 3D grid of $128\times128$
points in momentum space and $4096$ Matsubara frequencies. Actually,
many more cycles (almost twice) are needed at low doping and low temperatures.

Summarizing, within the NCA and the two-pole approximation for the
computation of $\chi_{\mu}(\mathbf{k},\omega)$ we have constructed
an analytical, completely self-consistent, scheme of calculation of
the electronic propagator $G_{cc}(\mathbf{k},\omega)=\mathcal{F}\langle T[c(i)c^{\dagger}(j)]\rangle$,
where dynamical contributions of the self-energy $\Sigma(\mathbf{k},\omega)$
are included. All the internal parameters are self-consistently calculated
by means of algebra constraints {[}cfr. (\ref{3.67})-(\ref{3.68})
and (\ref{3.71}){]}. No adjustable parameters or phenomenological
expressions are introduced.

\section{Results\label{sec:Results}}

In the following, we analyze some electronic properties by computing
the spectral function
\begin{equation}
A(\mathbf{k},\omega)=-\frac{1}{\pi}\Im[G_{cc}(\mathbf{k},\omega)]
\end{equation}
the momentum distribution function per spin 
\begin{equation}
n(\mathbf{k})=\int d\omega\, f_{\mathrm{F}}(\omega)A(\mathbf{k},\omega)
\end{equation}
and the density of states per spin 
\begin{equation}
N(\omega)=\frac{1}{(2\pi)^{2}}\int d^{2}kA(\mathbf{k},\omega)
\end{equation}
where $G_{cc}(\mathbf{k},\omega)=G_{11}(\mathbf{k},\omega)+G_{12}(\mathbf{k},\omega)+G_{21}(\mathbf{k},\omega)+G_{22}(\mathbf{k},\omega)$
is the electronic propagator and $f_{\mathrm{F}}(\omega)$ is the
Fermi function. We also study the electronic self-energy $\Sigma_{cc}(\mathbf{k},\omega)$,
which is defined through the equation 
\begin{equation}
G_{cc}(\mathbf{k},\omega)=\frac{1}{\omega-\epsilon_{0}(\mathbf{k})-\Sigma_{cc}(\mathbf{k},\omega)}
\end{equation}
where $\epsilon_{0}(\mathbf{k})=-\mu-4t\alpha(\mathbf{k})$ is the
non-interacting dispersion. Moreover, we define the quantity $r(\mathbf{k})=\epsilon_{0}(\mathbf{k})+\Sigma'_{cc}(\mathbf{k},\omega=0)$
that determines the Fermi surface locus in momentum space, $r(\mathbf{k})=0$,
in a Fermi liquid, i.e. when $\lim_{\omega\rightarrow0}\Sigma"_{cc}(\mathbf{k},\omega,T=0)\propto\omega^{2}$
and $\lim_{T\rightarrow0}\Sigma"_{cc}(\mathbf{k},\omega=0,T)\propto T^{2}$.
The actual Fermi surface (or its relic in a non-Fermi-liquid) is given
by the relative maxima of $A(\mathbf{k},\omega=0)$, which takes into
account, at the same time and on equal footing, both the real and
the imaginary parts of the self-energy and is directly related, within
the sudden approximation and forgetting any selection rules, to what
ARPES effectively measures.

Finally, the spin-spin correlation function $\left\langle n_{z}n_{z}^{\alpha}\right\rangle $,
the pole $\omega^{(3)}(\mathbf{k}=\mathbf{Q}=(\pi,\pi))$ of the spin-spin
propagator and the antiferromagnetic correlation length $\xi$. are
discussed. Usually, the latter is defined by supposing the following
asymptotic expression for the static susceptibility 
\begin{equation}
\lim_{\mathbf{k}\to\mathbf{Q}}\chi^{(3)}\left(\mathbf{k},0\right)=\frac{\chi^{(3)}\left(\mathbf{Q},0\right)}{1+\xi^{2}\left|\mathbf{k-Q}\right|^{2}}\label{xi}
\end{equation}
where $\chi^{(3)}\left(\mathbf{k},0\right)=-G^{(3)}\left(\mathbf{k},0\right)$.
It is worth noting that in our case (\ref{xi}) is not assumed, but
it exactly holds \cite{Avella_03a}.

\begin{figure}
\noindent \begin{centering}
\includegraphics[height=0.25\textheight]{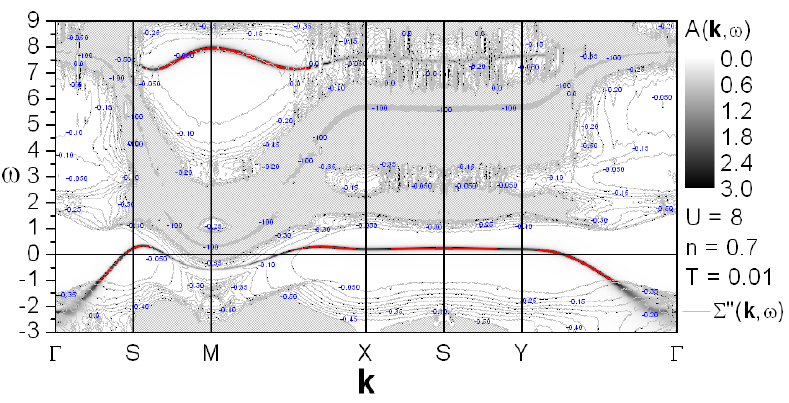}
\par\end{centering}

\noindent \begin{centering}
\includegraphics[height=0.25\textheight]{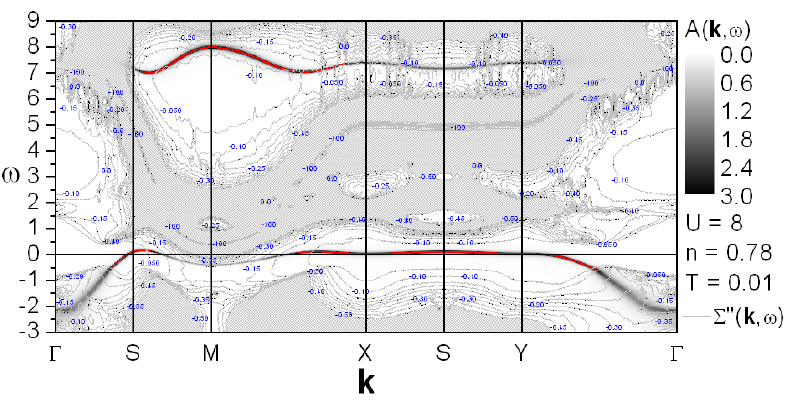}
\par\end{centering}

\noindent \begin{centering}
\includegraphics[height=0.25\textheight]{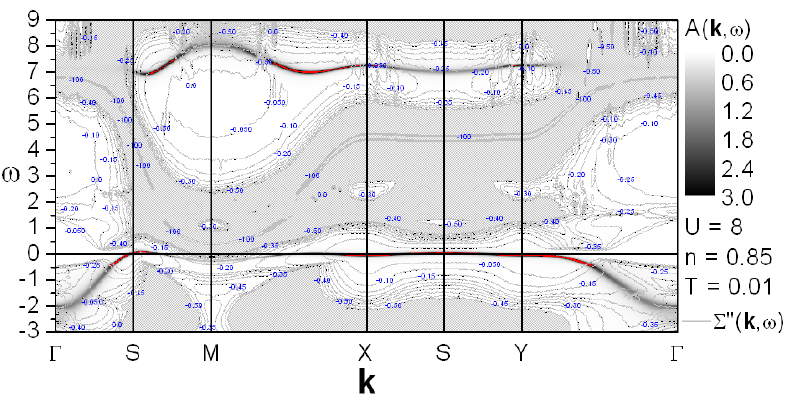}
\par\end{centering}

\noindent \centering{}\includegraphics[height=0.25\textheight]{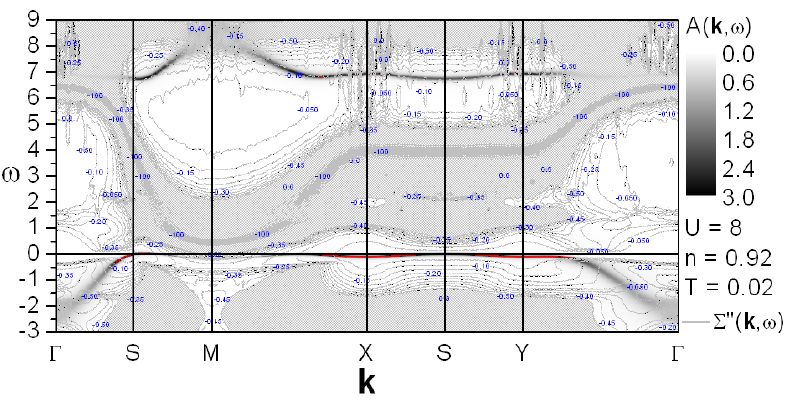}\protect\caption{(Color online) Spectral function $A(\mathbf{k},\omega)$ along the
principal directions ($\Gamma=(0,\,0)\to M=(\pi,\,\pi)$, $M\to X=(\pi,\,0)$,
$X\to Y=(0,\,\pi)$ and $Y\to\Gamma$) for $U=8$, $T=0.01$ and (top)
$n=0.70$, (middle top) $n=0.78$, (middle bottom) $n=0.85$ and (bottom)
$n=0.92$ ($T=0.02$).\label{fig9}}
\end{figure}

\begin{figure}
\noindent \begin{centering}
\includegraphics[height=0.25\textheight]{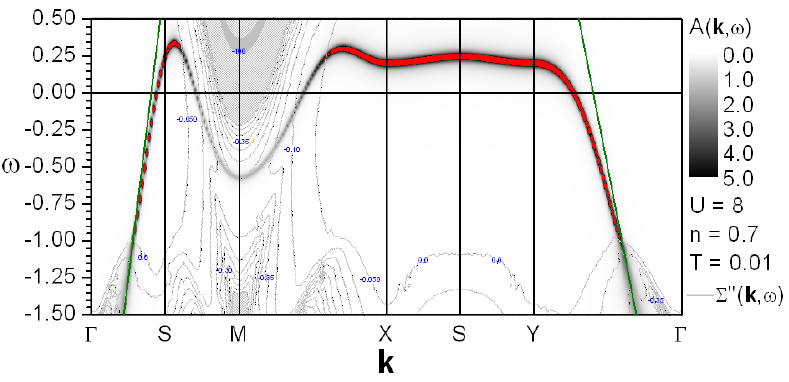} 
\par\end{centering}

\noindent \begin{centering}
\includegraphics[height=0.25\textheight]{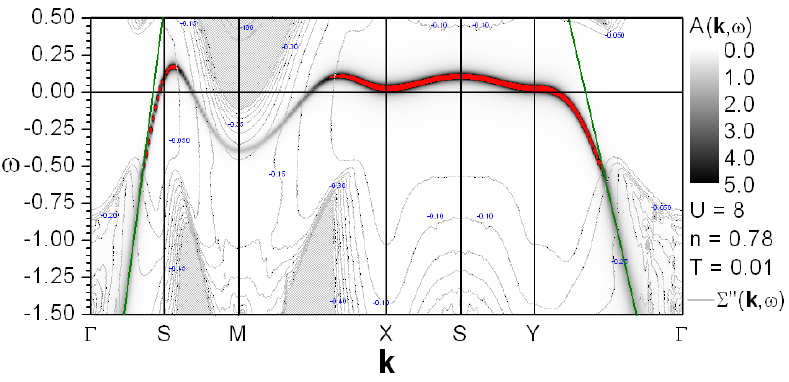}
\par\end{centering}

\noindent \begin{centering}
\includegraphics[height=0.25\textheight]{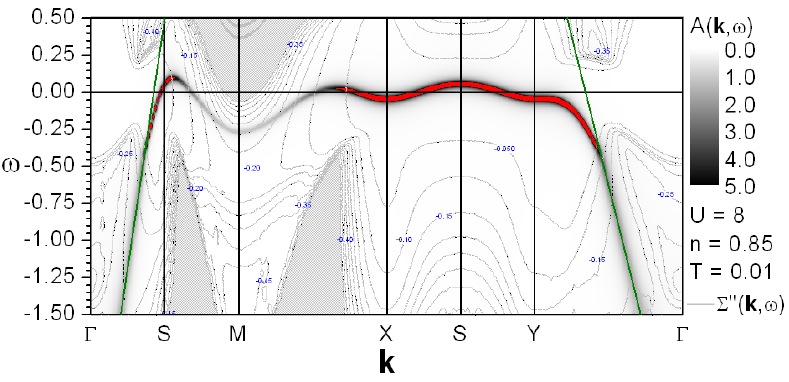}
\par\end{centering}

\noindent \centering{}\includegraphics[height=0.25\textheight]{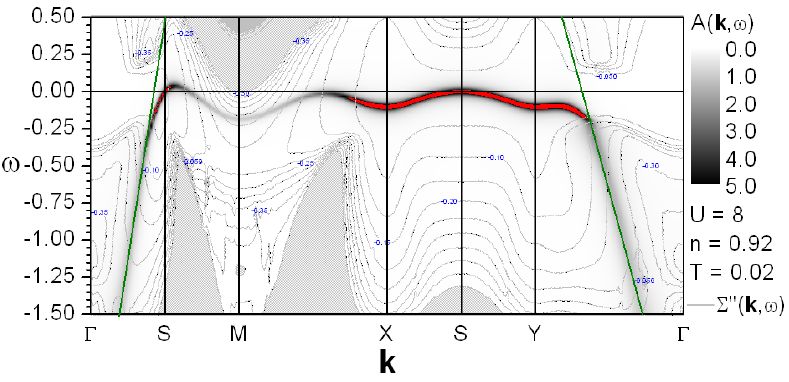}\protect\caption{(Color online) Spectral function $A(\mathbf{k},\omega)$ close to
the chemical potential ($\omega=0$) along the principal directions
($\Gamma=(0,\,0)\to M=(\pi,\,\pi)$, $M\to X=(\pi,\,0)$, $X\to Y=(0,\,\pi)$
and $Y\to\Gamma$) for $U=8$, $T=0.01$ and (top) $n=0.70$, (middle
top) $n=0.78$, (middle bottom) $n=0.85$ and (bottom) $n=0.92$ ($T=0.02$).\label{fig10}}
\end{figure}

\subsection{Spectral Function and Dispersion\label{sec:Dispersion}}

According to its overall relevance in the whole analysis performed
hereinafter, we first discuss the electronic dispersion of the model
under analysis or, better, its relic in a strongly correlated system.
In general, the dispersion and its more or less anomalous features
can be inferred by looking at the maxima of the spectral function
$A(\mathbf{k},\omega)$. In Figs.~\ref{fig9} and \ref{fig10}, the
spectral function is shown, in scale of grays (increasing from white
to black; red is for above-scale values), along the principal directions
($\Gamma=(0,\,0)\to M=(\pi,\,\pi)$, $M\to X=(\pi,\,0)$, $X\to Y=(0,\,\pi)$
and $Y\to\Gamma$) for $U=8$, $T=0.01$ and (top) $n=0.70$, (middle
top) $n=0.78$, (middle bottom) $n=0.85$ and (bottom) $n=0.92$ ($T=0.02$).
In Fig.~\ref{fig9}, the whole range of frequencies with finite values
of $A(\mathbf{k},\omega)$ is reported, while in Fig.~\ref{fig10}
a zoom in the proximity of the chemical potential is shown. The light
gray lines and uniform areas are labeled with the values of the imaginary
part of the self-energy $\Sigma''(\mathbf{k},\omega)$ and give one
of the most relevant keys to interpret the characteristics of the
dispersion. The dark green lines in Fig.~\ref{fig10} are just guides
to the eye and indicate the direction of the dispersion just \emph{before}
the visible kink separating the black and the red areas of the dispersion.

The red areas, as they mark the relative maxima of $A(\mathbf{k},\omega)$,
can be considered as the best possible estimates for the dispersion.
In a non- (or weakly-) interacting system, the dispersion would be
a single, continuos and (quite-)sharp line representing some function
$\varepsilon\left(\mathbf{k}\right)$ being the simple pole of $G_{cc}(\mathbf{k},\omega)$.
In this case (for a strongly correlated system), instead, we can clearly
see that the dispersion is well-defined (red areas) only in some of
the regions it crosses in the $(\mathbf{k},\omega)$ plane: the regions
where $\Sigma''(\mathbf{k},\omega)$ is zero or almost negligible.
In the crossed regions where $\Sigma''(\mathbf{k},\omega)$ is instead
finite, $A(\mathbf{k},\omega)$ obviously assumes very low values,
which would be extremely difficult (actually almost impossible) to
detect by ARPES. Accordingly, ARPES would report only the red areas
in the picture. This fact is fundamental to understand and describe
the experimental findings regarding the Fermi surface in the underdoped
regime, as they will discussed in the next section, and to reconcile
ARPES findings with those of quantum oscillations measurements.

The two Hubbard sub-bands, separated by a gap of the order $U$ and
with a reduced band-width of order $4t$, are clearly visible: the
lower one (LHB) is partly occupied as it is crossed by the chemical
potential, the upper one (UHB) is empty and very far from the chemical
potential. The lower sub-band systematically (for each value of doping)
loses significance close to $\Gamma$ and to $M$, although this effect
is more and more pronounced on reducing doping. In both cases (close
to $\Gamma$ and to $M$), $A(\mathbf{k},\omega)$ loses weight as
$\Sigma''(\mathbf{k},\omega)$ increases its own: this can be easily
understood if we recall that both $\chi_{0}\left(\mathbf{k},\omega\right)$
and $\chi_{3}\left(\mathbf{k},\omega\right)$ have a vanishing pole
at $\Gamma$ (due to hydrodynamics) and that $\chi_{3}\left(\mathbf{k},\omega\right)$
is strongly peaked at $M$ due to the strong antiferromagnetic correlations
present in the system (see Sec.~\ref{sec:Spin-dynamics}). The upper
sub-band, according to the complementary effect induced by the evident
shadow-bands appearance (signaled by the relative maxima of $\Sigma''(\mathbf{k},\omega)$
marked by $100$ and the dark gray area inside the gap) due again
to the strong antiferromagnetic correlations present in the system,
displays a well-defined dispersion at $M$, at least for high enough
doping, and practically no dispersion at all close to $\Gamma$. For
low doping, the great majority of the upper sub-band weight is simply
transferred to the lower sub-band. The growth, on reducing doping,
of the undefined-dispersion regions close to $\Gamma$ in the lower
sub-band cuts down its already reduced bandwidth of order $4t$ to
values of the order $J=\frac{4t^{2}}{U}=0.5t$, as one would expect
for the dispersion of few holes in a strong antiferromagnetic background.
The shape of the dispersion too is compatible with this scenario:
the sequence of minima and maxima is compatible, actually driven,
by the doubling of the Brillouin zone induced by the strong antiferromagnetic
correlations, as well as the dynamical generation of a $t'$ diagonal
hopping (absent in the Hamiltonian currently under study) clearly
signaled by the more and more, on reducing doping, pronounced warping
of the dispersion along the $X\to Y$ direction (more evident in Fig.~\ref{fig10}),
which would be perfectly flat otherwise (for $t'=0$).

Moving to the zooms (Fig.~\ref{fig10}), they show much more clearly:
the systematic reduction of the bandwidth on reducing doping, the
doubling of the zone, the systematic increase of the warping along
$X\to Y$ on reducing doping, the extreme flatness of the dispersion
at $X$ coming from both $\Gamma$ and $M$. This latter feature is
in very good agreement with quantum Monte Carlo calculations (see
\cite{Bulut_02} and references therein) as well as with ARPES experiments
\cite{Yoshida_01}, which report a similar behavior in the overdoped
region. Moreover, they show that, in contrast with the scenario for
a non- (or weakly-) interacting system, where the doubling of the
zone happens with the $X\to Y$ direction as \emph{pivot}, here the
doubling is confined to the region close to $M$. The lower sub-band
is completely filled at half-filling, while for an ordinary Slater
antiferromagnet the gap opens at half-filling just on top of the van-Hove
singularity along $X\to Y$. In addition, the effective, finite value
of $t'$ makes the dispersion maximum close to $S$ higher than the
one present along the $M\to X$ direction, opening the possibility
for the appearance of hole pockets close to $S$. Finally, it is now
evident that the warping of the dispersion along $X\to Y$ will also
induce the presence of two maxima in the density of states: one due
to the van-Hove singularities at $X$ and $Y$ and one due to the
maximum in the dispersion close to $S$. How deep is the dip between
these two maxima just depends on the number of available well-defined
(red) states in momentum present between these two values of frequency.
This will determine the appearance of a more or less pronounced pseudogap
in the density of states, but let us come back to this after having
analyzed the region close to $M$.

As a matter of fact, it is just the absence of spectral weight in
the region close to $M$ and, in particular and more surprisingly,
at the chemical potential (i.e. on the Fermi surface, in contradiction
with the Fermi-liquid picture), the main and more relevant result
of this analysis: it will determine almost all interesting and anomalous/unconventional
features of the single-particle properties of the model. The scenario
emerging from this analysis can be relevant not only for the understanding
of the physics of the Hubbard model and for the microscopical description
of the cuprate high-$T_{c}$ superconductors, but also for the drafting
of a general microscopic theory of strongly correlated systems. The
strong antiferromagnetic correlations (through $\Sigma''(\mathbf{k},\omega)$,
which mainly follows $\chi_{3}\left(\mathbf{k},\omega\right)$) cause
a significative and anomalous/unconventional loss of spectral weight
around $M$, which induces in turn: the deconstruction of the Fermi
surface (Sec.~\ref{sec:Spectral-Function-and}), the emergence of
momentum selective non-Fermi liquid features (Secs.~\ref{sec:Momentum-Distribution-Function}
and \ref{sec:Self-energy}), and the opening of a well-developed (deep)
pseudogap in the density of states (Sec.~\ref{sec:Density-of-States}).

Last, but not least, it is also remarkable the presence of kinks in
the dispersion in both the nodal ($\Gamma\to M$) and the antinodal
($X\to\Gamma$) directions, as highlighted by the dark green guidelines,
in qualitative agreement with some ARPES experiments \cite{Damascelli_03}.
Such a phenomenon clearly signals the coupling of the electrons to
a bosonic mode. In this scenario, the mode is clearly \emph{magnetic}
in nature. The frequency of the kink, with respect to the chemical
potential, reduces systematically and quite drastically on reducing
doping, following the behavior of the pole of $\chi_{3}\left(\mathbf{k},\omega\right)$
(see Sec.~\ref{sec:Spin-dynamics}). Combining the presence of kinks
and the strong reduction of spectral weight below them, we see the
appearance of waterfalls, in particular along the antinodal ($X\to\Gamma$)
direction, as found in some ARPES experiments \cite{Damascelli_03}.
Finally, the extension of the flat region in the dispersion around
the antinodal points ($X$ and $Y$) , i.e. at the van-Hove points,
increases systematically on decreasing doping. This clearly signals
the transfer of spectral weight from the Fermi surface, which is depleted
by the strong antiferromagnetic fluctuations, which are also responsible
for the remarkable flatness of the band edge.

Before moving to the next section, it is worth noticing that similar
results for the single-particle excitation spectrum (flat bands close
to $X$, weight transfer from the LHB to the UHB at $M$, splitting
of the band close to $X$, \dots ) were obtained within the self-consistent
projection operator method \cite{Kakehashi_04,Kakehashi_05}, the
operator projection method \cite{Onoda_01,Onoda_01a,Onoda_03} and
within a Mori-like approach by Plakida and coworkers \cite{Plakida_06,Plakida_10}.

\begin{figure}
\noindent \begin{centering}
\includegraphics[width=0.5\textwidth]{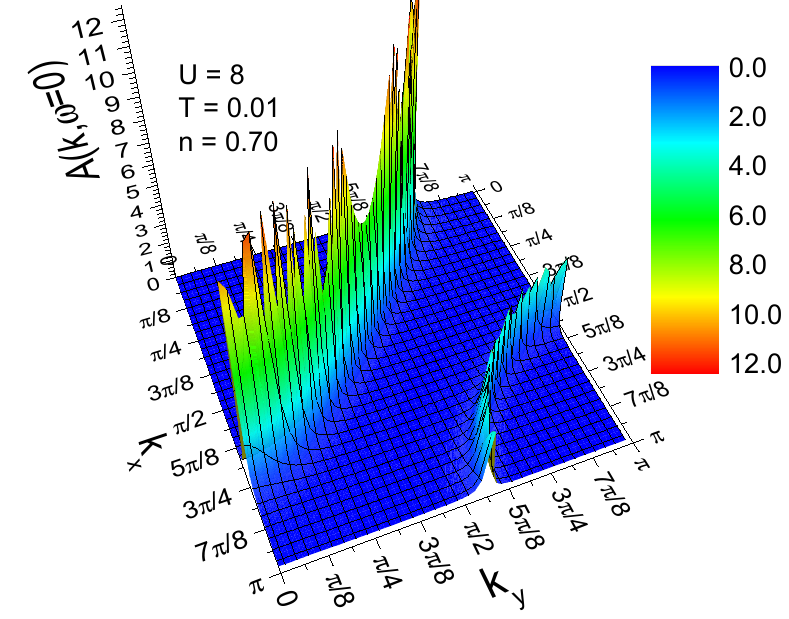}\hspace*{\fill}\includegraphics[width=0.5\textwidth]{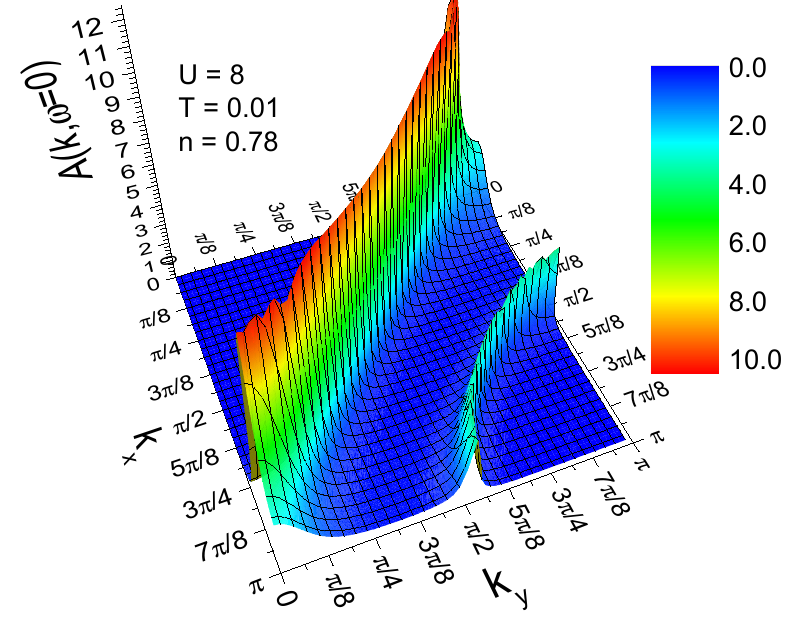}
\par\end{centering}

\noindent \centering{}\includegraphics[width=0.5\textwidth]{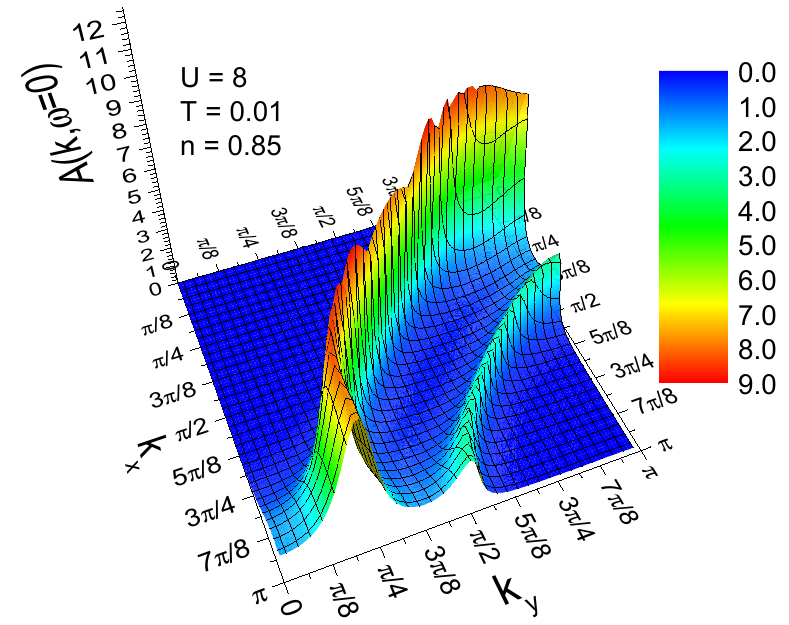}\hspace*{\fill}\includegraphics[width=0.5\textwidth]{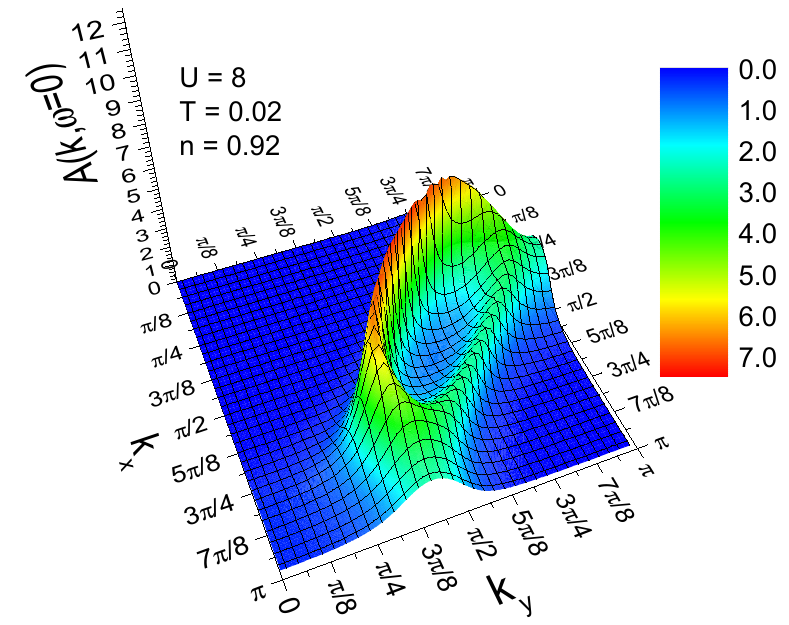}\protect\caption{(Color online) Spectral function at the chemical potential $A(\mathbf{k},\omega=0)$
as a function of momentum $\mathbf{k}$ for $U=8$, $T=0.01$ and
(top left) $n=0.70$, (top right) $n=0.78$, (bottom left) $n=0.85$
and (bottom right) $n=0.92$ ($T=0.02$).\label{fig1}}
\end{figure}

\begin{figure}
\noindent \begin{centering}
\includegraphics[width=0.5\textwidth]{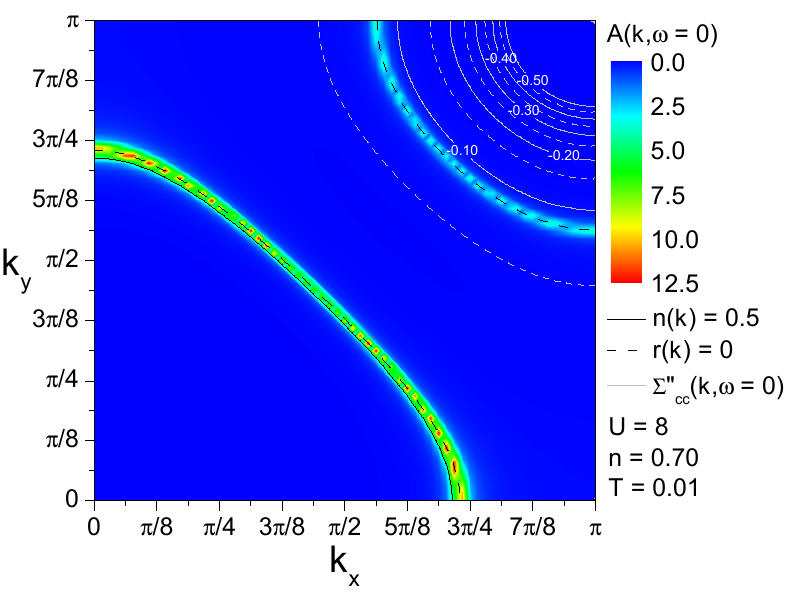}\hspace*{\fill}\includegraphics[width=0.5\textwidth]{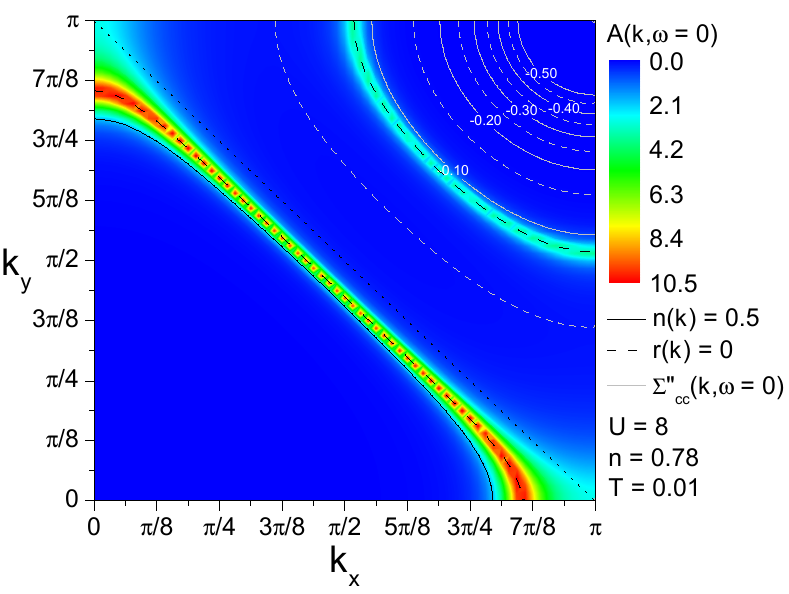}
\par\end{centering}

\noindent \centering{}\includegraphics[width=0.5\textwidth]{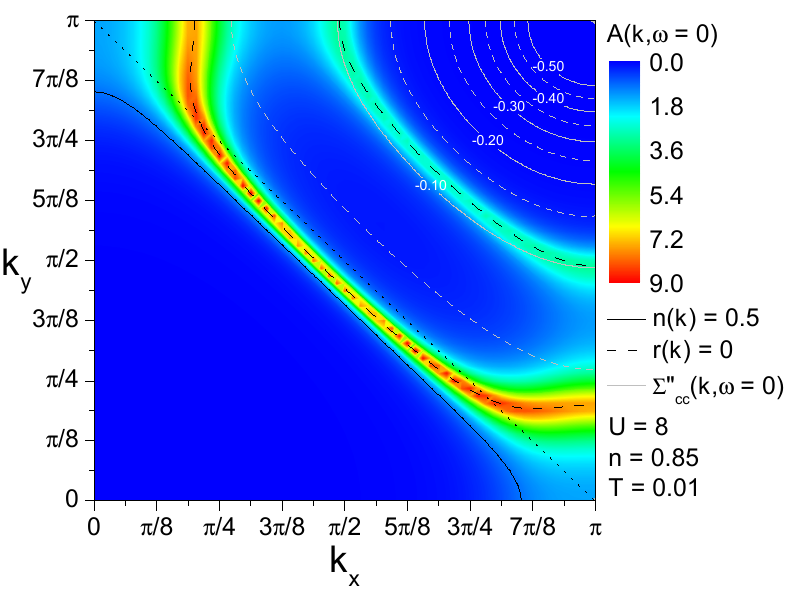}\hspace*{\fill}\includegraphics[width=0.5\textwidth]{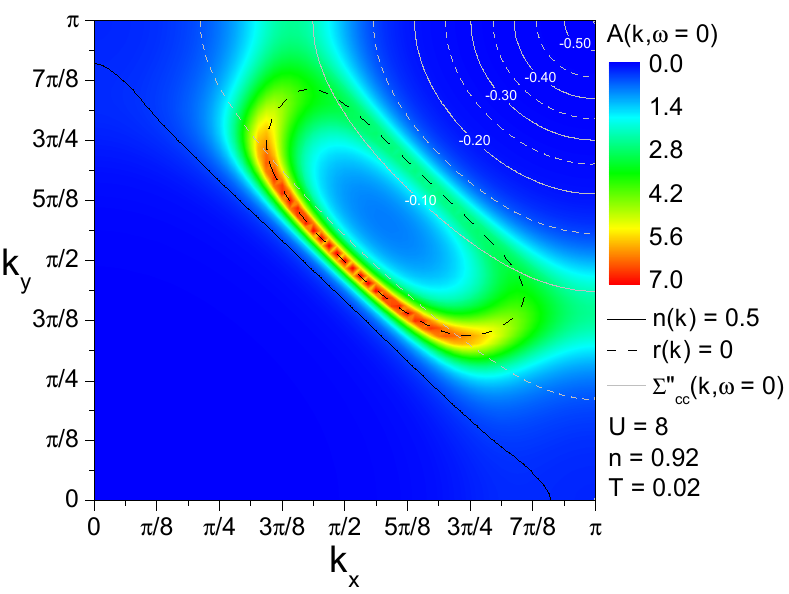}\protect\caption{(Color online) Spectral function at the chemical potential $A(\mathbf{k},\omega=0)$
as a function of momentum $\mathbf{k}$ for $U=8$, $T=0.01$ and
(top) $n=0.70$, (middle top) $n=0.78$, (middle bottom) $n=0.85$
and (bottom) $n=0.92$ ($T=0.02$). The solid line marks the locus
$n(\mathbf{k})=0.5$, the dashed line marks the locus $r(\mathbf{k})=0$,
the gray lines are labeled with the values of $\Sigma''_{cc}(\mathbf{k},\omega=0)$,
and the dotted line is a guide to the eye and marks the reduced (antiferromagnetic)
Brillouin zone.\label{fig2}}
\end{figure}

\subsection{Spectral Function and Fermi Surface\label{sec:Spectral-Function-and}}

Focusing on the value of the spectral function at the chemical potential,
$A(\mathbf{k},\omega=0)$, we can discuss the closest concept to Fermi
surface available in a strongly correlated system. In Fig.~\ref{fig1},
$A(\mathbf{k},\omega=0)$ is plotted as a function of the momentum
$\mathbf{k}$ in a quarter of the Brillouin zone for $U=8$ and four
different couples of values of temperature and filling: (top left)
$n=0.7$ and $T=0.01$, (top right) $n=0.78$ and $T=0.01$ , (bottom
left) $n=0.85$ and $T=0.01$ and (bottom right) $n=0.92$ and $T=0.02$.
The Fermi surface, in agreement with the interpretation of the ARPES
measurements within the sudden approximation, can be defined as the
locus in momentum space of the relative maxima of $A(\mathbf{k},\omega=0)$.
Such a definition opens up the possibility to explain ARPES measurements,
but also to go beyond them and their finite instrumental resolution
and sensitivity with the aim at filling the gap with other kind of
measurements, in particular quantum oscillations ones, which seems
to report results in disagreement, up to dichotomy in some cases,
with the scenario depicted by ARPES. 

\begin{figure}
\noindent \centering{}\mbox{}\hspace*{\fill}\includegraphics[height=6cm]{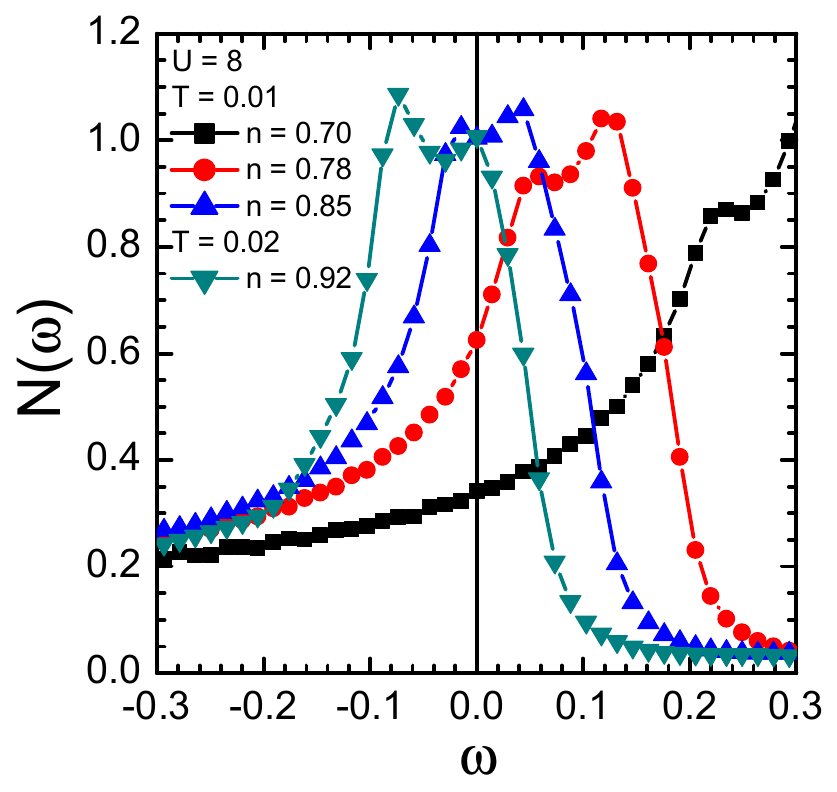}\hspace*{\fill}\includegraphics[height=6cm]{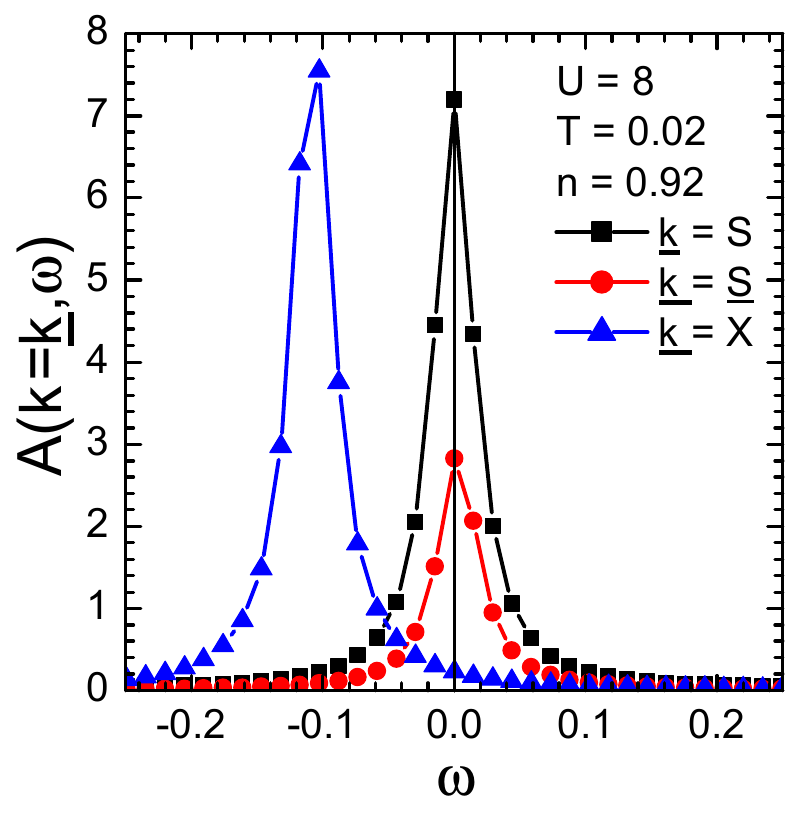}\hspace*{\fill}\mbox{}\protect\caption{(Color online) (left) Density of states $N(\omega)$ as a function
of frequency $\omega$ for $U=8$, (black squares) $n=0.7$ and $T=0.01$,
(red circles) $n=0.78$ and $T=0.01$, (blue up triangles) $n=0.85$
and $T=0.01$, and (green down triangles) $n=0.92$ and $T=0.02$.
(right) Spectral function in proximity of the chemical potential $A(\mathbf{\underline{k}},\omega\sim0)$
at (black squares) $\mathbf{\underline{k}}=S=(\pi/2,\pi/2)$, (red
circles) $\underline{S}$ (in the text), and (blue triangles) $X=(\pi,\,0)$
for $U=8$, $n=0.92$ and $T=0.02$.\label{fig8}}
\end{figure}

For each value of the filling reported, we can easily distinguish
two walls/arcs; for $n=0.92$, they somewhat join. First, let us focus
on the arc with the larger (by far) intensities; given the current
sensitivities, this is the only one, among the two, possibly visible
to ARPES. At $n=0.7$ (see Fig.~\ref{fig1} (top left panel)), the
3D perspective allows to better appreciate the difference in the intensities
of the relative maxima of $A(\mathbf{k},\omega=0)$ between the region
close to the main diagonal ($M\to X$), where the signal is weaker,
and the regions close to the main axes ($\Gamma\to X$ and $\Gamma\to Y$),
where the signal is stronger; this behavior has been also reported
by ARPES experiments \cite{Yoshida_01,Damascelli_03} as well as the
electron-like nature of the Fermi surface \cite{Yoshida_01}. On decreasing
doping, this trend reverses, passing through $n=0.78$, where the
intensities almost match, and up to $n=0.92$, where the region in
proximity of $S$ is the only one with an appreciable signal. The
less-intense (by far) arc, reported in Ref.~\cite{Plakida_06} too,
is the relic of a shadow band, as can be clearly seen in Fig.~\ref{fig10},
and, consequently, never changes its curvature, in contrast to what
happens to the other arc, which is subject to the crossing of the
van Hove singularity ($n\cong0.82$) instead. Although the ratio between
the maximum values of the intensities at the two arcs never goes below
two (see Fig~\ref{fig8} (right panel)), there is an evident decrease
of the maximum value of the intensity at the larger-intensity arc
on decreasing doping, which clearly signals an overall increase of
the intensity and/or of the effectiveness (in terms of capability
to affect the relevant quasi-particles, which are those at the Fermi
surface) of the correlations.

Moving to a 2D perspective (see Fig.~\ref{fig2}), we can add three
ingredients to our discussion that can help us better understanding
the evolution with doping of the Fermi surface: (i) the $n(\mathbf{k})=0.5$
locus (solid line), i.e. the Fermi surface if the system would be
non-interacting; (ii) the $r(\mathbf{k})=0$ locus (dashed line),
i.e. the Fermi surface if the system would be a Fermi liquid or somewhat
close to it conceptually; (iii) the values (grey lines and labels)
of the imaginary part of the self-energy at the chemical potential
$\Sigma''_{cc}(\mathbf{k},\omega=0)$ (notice that $T\neq0$). Combining
these three ingredients with the positions and intensities of the
the relative maxima of $A(\mathbf{k},\omega=0)$, we can try to better
understand what these latter signify and to classify the behavior
of the system on changing doping. At $n=0.7$ (see Fig.~\ref{fig2}
(top left panel)), the positions of the two arcs are exactly matching
$r(\mathbf{k})=0$ lines; this will stay valid at each value of the
filling reported, with a fine, but very relevant, distinction at $n=0.92$.
This occurrence makes our definition of Fermi surface robust, but
also versatile as it permits to go beyond Fermi liquid picture without
contradicting this latter. The almost perfect coincidence, for the
higher value of doping reported, $n=0.7$, of the $n(\mathbf{k})=0.5$
line with the larger-intensity arc clearly asserts that we are dealing
with a very-weakly-interacting Fermi metal. $\Sigma''_{cc}(\mathbf{k},\omega=0)$
is quite large close to $M$ (see Fig.~\ref{fig10} (top panel))
and eats up the weight of the second arc, which becomes a \emph{ghost}
band more than a \emph{shadow} one. The antiferromagnetic correlations
are definitely finite (see Sec.~\ref{sec:Spin-dynamics}) and, consequently,
lead to the doubling of the zone, but not strong enough to affect
the behavior of the ordinary quasi-particles safely living at the
ordinary Fermi surface. Decreasing the doping, we can witness a first
topological transition from a Fermi surface closed around $\Gamma$
(electron like - hole like in cuprates language) to a Fermi surface
closed around $M$ (hole like - electron like in cuprates language)
at $n\cong0.82$, where the chemical potential crosses the van Hove
singularity (see Fig.~\ref{fig10}). The chemical potential presents
an inflection point at this doping (not shown), which allowed us to
determine its value with great accuracy. In proximity of the antinodal
points ($X$ and $Y$), a net discrepancy between the position of
the relative maxima of $A(\mathbf{k},\omega=0)$ and the locus $n(\mathbf{k})=0.5$
becomes more and more evident on decreasing doping (see Fig.~\ref{fig2}
(top right and bottom left panels)). This occurrence does not only
allows the topological transition, which is absent for the $n(\mathbf{k})=0.5$
locus that reaches the anti-diagonal ($X\to Y$) at half-filling in
agreement with the Luttinger theorem, but it also accounts for the
apparent broadening of the relative maxima of $A(\mathbf{k},\omega=0)$
close to the anti-nodal points ($X$ and $Y$). The broadening is
due to the small, but finite, value of $\Sigma''_{cc}(\mathbf{k},\omega=0)$
in those momentum regions (see Fig.~\ref{fig10} (middle panels)),
signaling the net increase of the correlation strength and, accordingly,
the impossibility to describe the system in this regime as a conventional
non- (or weakly-) interacting system within a Fermi-liquid scenario
or its ordinary extensions for ordered phases. What is really interesting
and goes beyond the actual problem under analysis (cuprates - 2D Hubbard
model), is the emergence of such features only in well defined regions
in momentum space. This selectiveness in momentum is quite a new feature
in condensed matter physics and its understanding and description
require quite new theoretical approaches. In this system, almost independently
from their effective strength, the correlations play a so fundamental
role to come to shape and determine \emph{qualitatively} the response
of the system. Accordingly, any attempt to treat correlations without
taking into account the level of entanglement between all degrees
of freedom present in the system is bound to fail or at least to miss
the most relevant features.

Let us come to the most interesting result. At $n=0.92$, the relative
maxima of $A(\mathbf{k},\omega=0)$ detach also from the $r(\mathbf{k})=0$
locus, at least partially, opening a completely new scenario. $r(\mathbf{k})=0$
defines a pocket (close, but absolutely not identical - read below,
to that of an antiferromagnet), while the relative maxima of $A(\mathbf{k},\omega=0)$
feature the very same pocket together with quite broad, but still
well-defined, wings closing with one half of the pocket (the most
intense one) what can be safely considered the relic of a large Fermi
surface. This is the second and most surprising topological transition
occurring to the actual Fermi surface: the two arcs, clearly visible
for all other values of the filling, join and instead of closing just
a pocket, as one would expect on the basis of the conventional theory
for an antiferromagnet - here mimed by $r(\mathbf{k})=0$ locus, develop
(or keep) a completely independent branch. The actual Fermi surface
is neither a pocket nor a large Fermi surface; for a more expressive
representation see Fig.~\ref{fig1} (bottom right panel). This very
unexpected result can be connected to the dichotomy between those
experiments (e.g. ARPES) pointing to a small and those ones (e.g.
quantum oscillations) pointing to a large Fermi surface. This result
can be understood by looking once more at the dispersion for this
value of the filling (see Fig.~\ref{fig10} (bottom panel)): the
difference in height, induced by the effective finite value of $t'$,
which increases with decreasing doping, between the two highest maxima
in the dispersion - one close to $S$ and the other along the $M\to X$
direction - makes the latter to cross the chemical potential for larger
values of the doping than the first, but given the significative broadening
of the dispersion, even when the center mass of the second leaves
the Fermi surface (disappearing from $r(\mathbf{k})=0$ locus), its
shoulders are still active and well-identifiable at the chemical potential
- that is on the actual Fermi surface.

The pocket too is far from being \emph{conventional}. It is clearly
evident in Fig.~\ref{fig2} (bottom right panel) that there are two
distinct halves of the pocket: one with very high intensity pinned
at $S$ (again the only possibly visible to ARPES) and another with
very low intensity (visible only to \emph{theoreticians} and some
quantum oscillations experiments). This is our interpretation for
the Fermi arcs reported by many ARPES experiments\cite{Damascelli_03,Shen_05}
and unaccountable for any ordinary theory relaying on the Fermi liquid
picture, although modified by the presence of an incipient spin or
charge ordering. Obviously, looking only at the Fermi arc (as ARPES
is forced to do), the Fermi surface looks ill defined as it does not
enclose a definite region of momentum space, but having access also
to the other half of the pocket, such problem is greatly alleviated.
The point is that the antiferromagnetic fluctuations are so strong
to destroy the coherence of the quasi-particles in that region of
momentum space as similarly reported within the DMFT$+\Sigma$ approach
\cite{Sadovskii_01,Sadovskii_05,Kuchinskii_05,Kuchinskii_06,Kuchinskii_06a}
and a Mori-like approach by Plakida and coworkers \cite{Plakida_06,Plakida_10}.
Moreover, we will see that (see Sec.~\ref{sec:Self-energy}), the
\emph{phantom} half is not simply lower in intensity because it belongs
to a shadow band depleted by a finite imaginary part of the self-energy
(see Fig.~\ref{fig2} (bottom right panel)), but that it lives in
a region of momentum where the imaginary part of the self-energy shows
clear signs of non-Fermi liquid behavior. The lack of next-nearest
hopping terms (i.e., $t'$ and $t''$) in the chosen Hamiltonian (\ref{3.1}),
although they are evidently generated dynamically, does not allow
us to perform a quantitative comparison between our results and the
experimental ones, which refers to a specific material characterized
by a specific set of hoppings. This also explains why our Fermi arc
is pinned at $S$ and occupies the outer reduced Brillouin zone, while
many experimental results report a Fermi arc occupying the inner reduced
Brillouin zone. Actually, the pinning (with respect to doping within
the underdoped region) of the \emph{center of mass} of the ARPES-visible
Fermi arc has been reported also from ARPES experiments\cite{Koitzsch_04}.

\subsection{Density of States and Pseudogap\label{sec:Density-of-States}}

The other main issue in the underdoped regime of cuprates superconductors
is the presence of a quite strong depletion in the electronic density
of states, known as pseudogap. In Fig.~\ref{fig8} (left panel),
we report the density of states $N(\omega)$ for $U=8$ and four couples
of values of filling and temperature: $n=0.7$ and $T=0.01$, $n=0.78$
and $T=0.01$, $n=0.85$ and $T=0.01$, and $n=0.92$ and $T=0.02$,
in the frequency region in proximity of the chemical potential. As
a reference, we also report, in Fig.~\ref{fig8} (right panel), the
spectral function in proximity of the chemical potential $A(\mathbf{\underline{k}},\omega\sim0)$
at $\mathbf{\underline{k}}=S=(\pi/2,\pi/2)$, $\mathbf{\underline{k}}=\underline{S}$
which lies where the \emph{phantom} half of the pocket touches the
main diagonal $\Gamma\to M$ (i.e. where the dispersion cuts the main
diagonal $\Gamma\to M$ closer to $M$), and $\mathbf{\underline{k}}=X=(\pi,\,0)$
for $U=8$, $n=0.92$ and $T=0.02$. As it can be clearly seen in
Fig.~\ref{fig8} (left panel), the density of states present two
maxima separated by a dip, which plays the role of pseudogap in this
scenario. Its presence is due to the warping in the dispersion along
the $X\to Y$ direction (see Fig.~\ref{fig10}), which induces the
presence of the two maxima (one due to the van-Hove singularity at
$X$ and one due to the maximum in the dispersion close to $S$ -
see Fig.~\ref{fig8} (right panel)), and to the loss of states, within
this window in frequency, in the region in momentum close to $M$,
as discussed in detail in the previous sections. As a measure of how
much weight is lost because of the finite value of the imaginary part
of the self-energy in the region in momentum close to $M$, one can
look at the striking difference between the value of $A(\underline{S},\omega=0)$
in comparison to that of $A(S,\omega=0)$, see Fig.~\ref{fig8} (right
panel). On reducing the doping, there is an evident transfer of spectral
weight between the two maxima; in particular, the weight is transferred
from the top of the dispersion close to $S$ to the antinodal point
$X$, where the van Hove singularity resides. At the lowest doping
($n=0.92$), a well developed pseudogap is visible below the chemical
potential and will clearly affect all measurable properties of the
system. For this doping, we do not observe any divergence of $\Sigma'_{cc}(\mathbf{k},\omega=0)$
in contrast to what reported in Ref.~\cite{Stanescu_05} where this
feature is presented as the ultimate reason for the opening of the
pseudogap. In our scenario, the pseudogap is just the result of the
transfer of weight from the single-particle density of states to the
two-particle one related to the (antiferro)magnetic excitations developing
in the system on decreasing doping at low temperatures (see Sec.~\ref{sec:Spin-dynamics}).

It is worth mentioning that an analogous doping behavior of the pseudogap
has been found by the DMFT$+\Sigma$ approach \cite{Sadovskii_01,Sadovskii_05,Kuchinskii_05,Kuchinskii_06,Kuchinskii_06a},
a Mori-like approach by Plakida and coworkers \cite{Plakida_06,Plakida_10}
and the cluster perturbation theory \cite{Senechal_00,Senechal_04,Tremblay_06}.

\begin{figure}[p]
\noindent \centering{}\mbox{}\hspace*{\fill}\includegraphics[height=5cm]{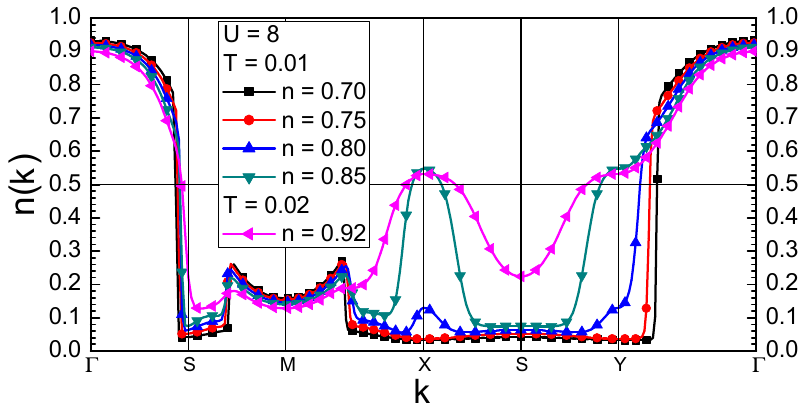}\hspace*{\fill}\includegraphics[height=5cm]{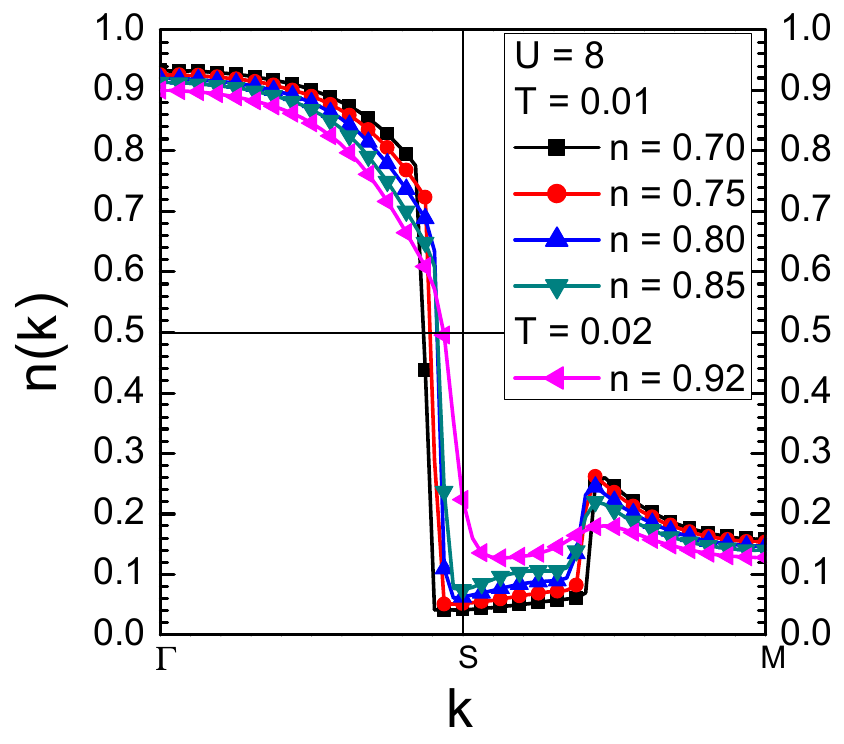}\hspace*{\fill}\mbox{}\protect\caption{(Color online) Momentum distribution function $n(\mathbf{k})$ for
$U=8$ and $n=0.7$, $0.75$, $0.8$, $n=0.85$ ($T=0.01$), and $n=0.92$
($T=0.02$): (left) along the principal directions ($\Gamma=(0,\,0)\to M=(\pi,\,\pi)$,
$M\to X=(\pi,\,0)$, $X\to Y=(0,\,\pi)$ and $Y\to\Gamma$); (right)
along the principal diagonal ($\Gamma\to M$).\label{fig4}}
\end{figure}

\begin{figure}[p]
\noindent \centering{}\includegraphics[width=0.5\textwidth]{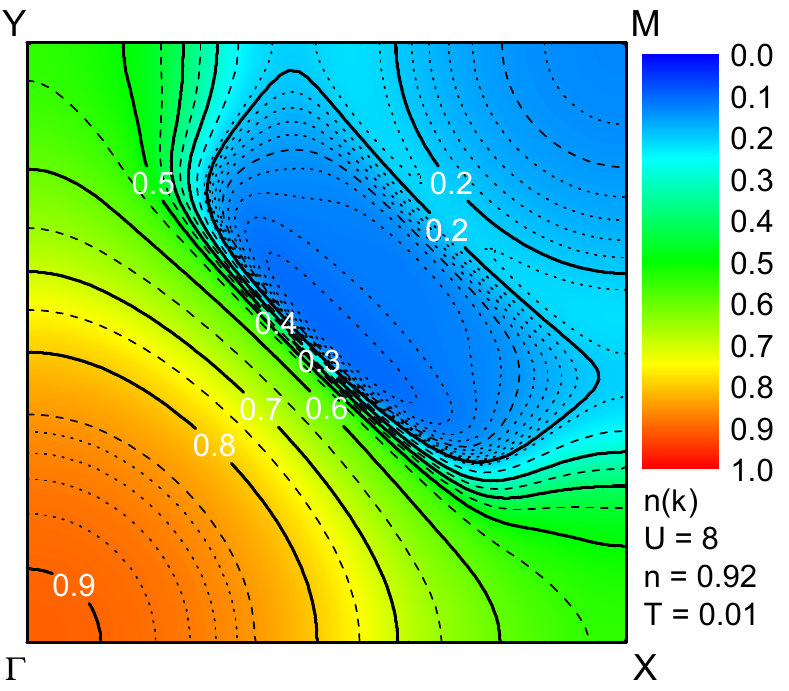}
\protect\caption{(Color online) The momentum distribution function $n(\mathbf{k})$
for $n=0.92$, $T=0.01$ and $U=8$.\label{fig6}}
\end{figure}

\begin{figure}[p]
\noindent \centering{}\mbox{}\hspace*{\fill}\includegraphics[height=5cm]{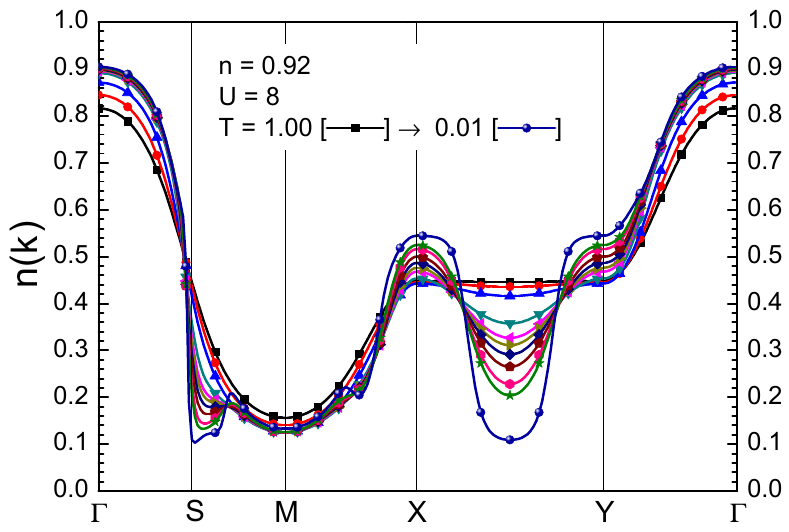}\hspace*{\fill}\includegraphics[height=5cm]{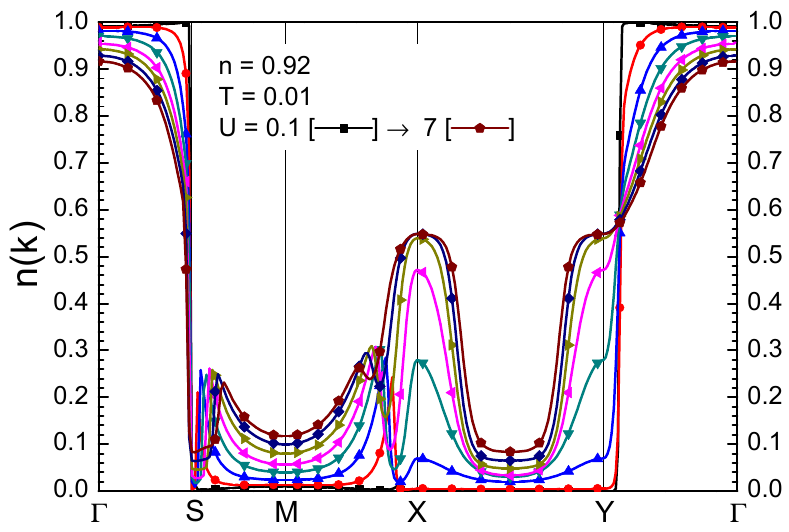}\hspace*{\fill}\mbox{}\protect\caption{(Color online) The momentum distribution function $n(\mathbf{k})$
along the main directions ($\Gamma=(0,0)$ $\to$ $M=(\pi,\pi)$ $\to$
$X=(\pi,0)$ $\to$ $Y=(0,\pi)$ $\to$ $\Gamma=(0,0)$) for different
values of temperature $T$ (left) and on-site Coulomb repulsion $U$
(right) at $n=0.92$.\label{fig5}}
\end{figure}

\subsection{Momentum Distribution Function\label{sec:Momentum-Distribution-Function}}

To analyze a possible crossover from a Fermi liquid to a non-Fermi
liquid behavior in certain regions of momentum space, at small dopings
and low temperatures, and to better characterize the pocket forming
on the Fermi surface at the lowest reported doping, we study the electronic
momentum distribution function $n(\mathbf{k})$ per spin along the
principal directions ($\Gamma\to M$, $M\to X$, $X\to Y$ and $Y\to\Gamma$)
and report it in Fig.~\ref{fig4}, for $U=8$ and $n=0.7$, $0.75$,
$0.8$, and $0.85$ ($T=0.01$), and for $n=0.92$ ($T=0.02$). The
right panel in the figure reports a zoom along the main diagonal ($\Gamma\to M$).
At the highest studied doping ($n=0.7$), $n(\mathbf{k})$ shows the
usual features of a quasi non-interacting system, except for one single,
but very important feature: the dip along the main diagonal ($\Gamma\to M$)
signaling the presence of a shadow band due to the weak, but anyway
finite, antiferromagnetic correlations (see Sec.~\ref{sec:Spin-dynamics}),
as already discussed many times hereinbefore. The quite small height
of the \emph{secondary} jumps along $S\to M$ and $M\to X$ directions
(with respect to the height of the \emph{main} jumps along $\Gamma\to S$
and $Y\to\Gamma$ directions) gives a measure of the relevance of
this feature in the overall picture: not really much relevant, except
at $n=0.92$ where it changes qualitatively. On increasing filling
(reducing doping), the features related to the shadow band do not
change much their positions and intensity, while the features related
to the ordinary band change their positions as expected in order to
\emph{accommodate} (i.e. to activate states in momentum for) the increasing
number of particles. Finally, at $n=0.92$, the two sets of features
close a pocket. At this final stage, what is very relevant, as it
is very unconventional, is the evident and remarkable difference in
behavior between the \emph{main} jump along the $\Gamma\to S$ direction
and the \emph{secondary} jump along the $S\to M$ direction (see Fig.~\ref{fig4}
(right panel)): the former stays quite sharp (just a bit skewed by
the slightly higher temperature) as the Fermi liquid theory requires,
the latter instead loses completely its sharpness, much more than
what would be reasonable because of the finite value of the temperature
as it can be deduced by the comparison with the behavior of the \emph{main}
jump. Such a strong qualitative modification is the evidence of a
non-Fermi-liquid-like kind of behavior, but only combining this occurrence
with a detailed study of the frequency and temperature dependence
of the imaginary part of the self-energy in the very same region of
momentum space (see Sec.~\ref{sec:Self-energy}), we will be able
to make a definitive statement about this.

In Fig.~\ref{fig6}, we try to summarize the scenario in the extreme
case ($n=0.92$, $T=0.01$ and $U=8$), where all the anomalous features
are present and well formed, by reporting the full 2D scan of the
momentum distribution function $n(\mathbf{k})$ in a quarter of the
Brillouin zone. We can clearly see now the pocket with its center
along the main diagonal and the lower border touching the border of
the magnetic zone at exactly $S=(\pi/2,\pi/2)$. Actually, the 2D
prospective makes more evident that there is a second underlying Fermi
surface that corresponds to the ordinary paramagnetic one for this
filling $n=0.92$ (large and hole-like) and touching the border of
the zone between $M=(\pi,\pi)$ and $X=(\pi,0)$ ($Y=(0,\pi)$). This
corresponds to the very small jump in Fig.~\ref{fig5} (left panel)
along the same direction. It is worth mentioning that a similar behavior
of the momentum distribution function has been found by means of a
Mori-like approach by Plakida and coworkers \cite{Plakida_06,Plakida_10}.

In Fig.~\ref{fig5}, we study the dependence of the momentum distribution
function $n(\mathbf{k})$ on the temperature $T$ (left panel) and
the on-site Coulomb repulsion $U$ (right panel) by keeping the filling
$n$ fixed at the most interesting value: $0.92$. At high temperatures
(in particular, down to $T=0.4$), the behavior of $n(\mathbf{k})$
is that of a weakly correlated paramagnet (no pocket along the $S\to M$
direction, no warping along the $X\to Y$ direction). For lower temperatures,
the pocket develops along the $S\to M$ direction and the signal along
the $X\to Y$ direction is no more constant, signaling the dynamical
generation of a diagonal hopping term $t'$, connecting same-spin
sites in a newly developed antiferromagnetic background unwilling
to be disturbed. In fact, such a behavior is what one expects when
quite strong magnetic fluctuations develop in the system and corresponds
to a well defined tendency towards an antiferromagnetic phase (see
Sec.~\ref{sec:Spin-dynamics}). The $M$ point becomes another minimum
in the dispersion in competition with $\Gamma$ and the dispersion
should feature a maximum between them in correspondence to the center
of the pocket. The whole bending of the dispersion confines the van
Hove singularity below the Fermi surface in a \emph{open} pocket (it
closes out of the actually chosen Brillouin zone) visible in the momentum
distribution as a new quite broad maximum at $X$ and $Y$. The dependence
on $U$ shows that for $n=0.92$ and $T=0.01$, our solution presents
quite strong antiferromagnetic fluctuations for every finite value
of the Coulomb repulsion, although the two kind of pockets discussed
just above are not well formed for values of $U$ less than $U=3\div4$.

\begin{figure}
\noindent \begin{centering}
\mbox{}\hspace*{\fill}\includegraphics[height=6cm]{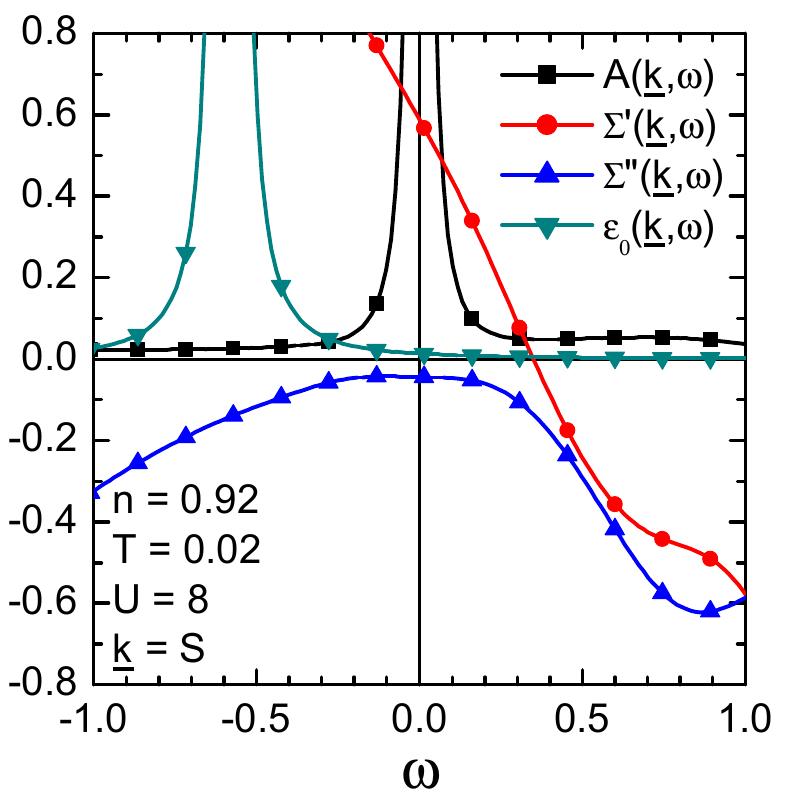}\hspace*{\fill}\includegraphics[height=6cm]{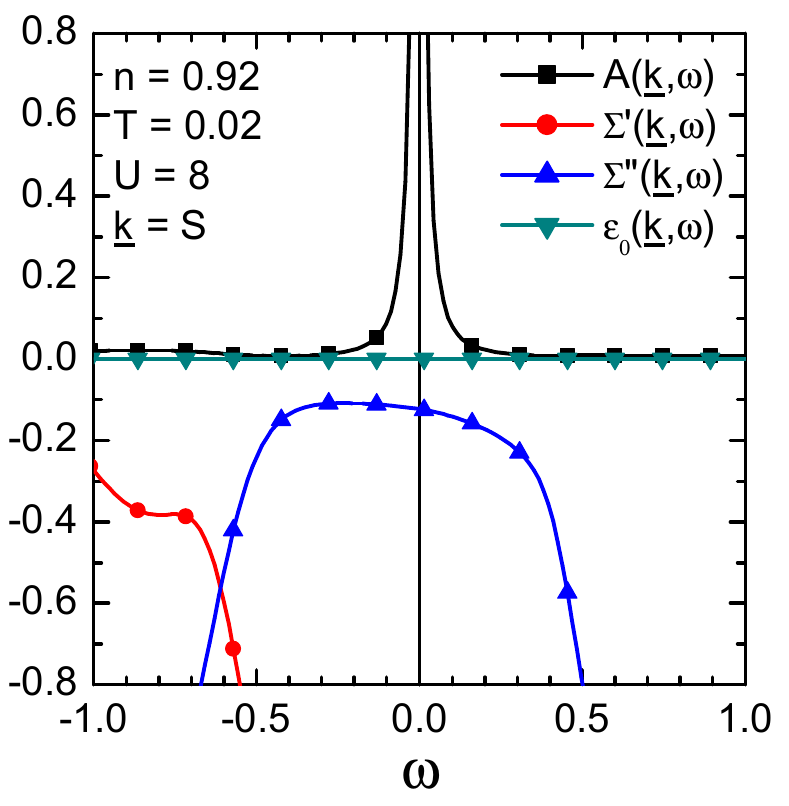}\hspace*{\fill}\mbox{}
\par\end{centering}

\noindent \centering{}\mbox{}\hspace*{\fill}\includegraphics[height=6cm]{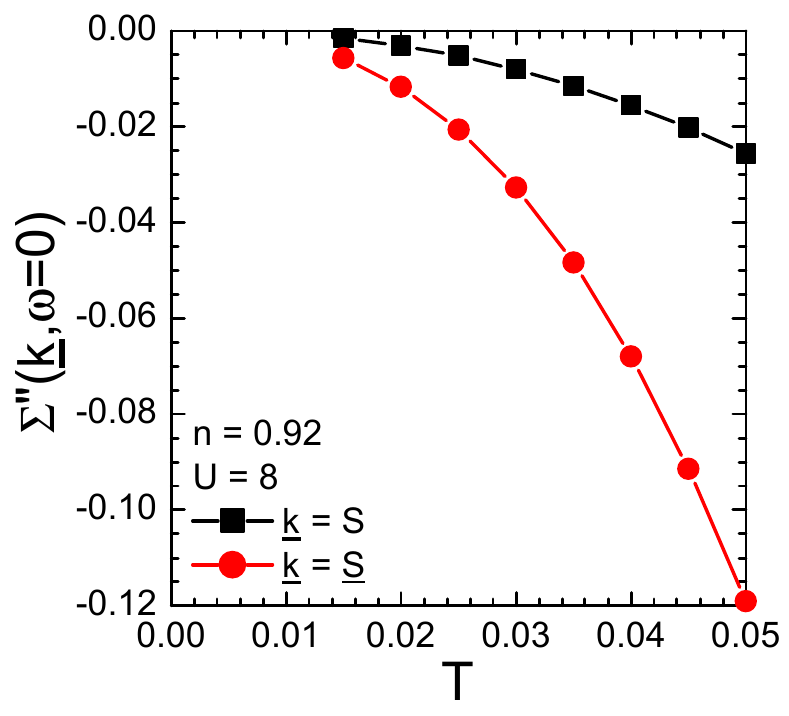}\hspace*{\fill}\includegraphics[height=6cm]{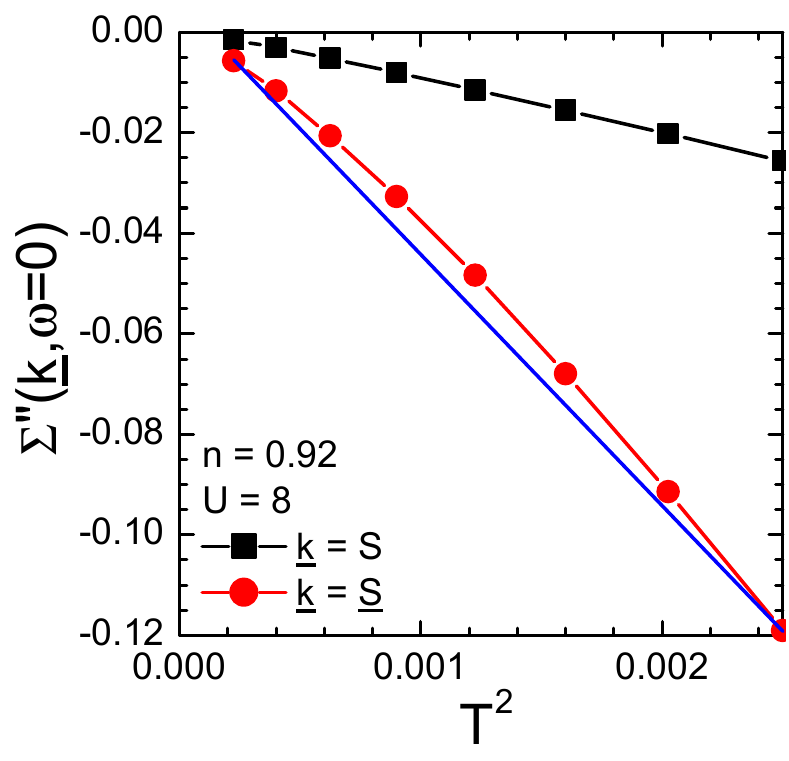}\hspace*{\fill}\mbox{}\protect\caption{(Color online) Spectral density function $A\left(\underline{\mathbf{k}},\omega\right)$,
real ($'$) and imaginary ($''$) part of the self-energy $\Sigma\left(\underline{\mathbf{k}},\omega\right)$,
non-interacting dispersion $\varepsilon_{0}\left(\underline{\mathbf{k}},\omega\right)$
as functions of frequency at (top left) $\underline{\mathbf{k}}=S$
and (top right) $\underline{\mathbf{k}}=\underline{S}$ for $n=0.92$,
$T=0.02$ and $U=8$. (bottom) Imaginary part of the self-energy $\Sigma''\left(\underline{\mathbf{k}},\omega=0\right)$
as function of temperature at (squares) $\mathbf{k}=S$ and (circles)
$\mathbf{k}=\underline{S}$ for $n=0.92$ and $U=8$. The blue line
is just a guide to the eye.\label{fig7}}
\end{figure}

\subsection{Self-energy\label{sec:Self-energy}}

To obtain clear-cut pieces of information about the lifetime of the
quasi-particles generated by the very strong interactions within this
scenario, but even more to understand if it is still reasonable or
not to discuss in terms of quasi-particles at all (i.e. if this Fermi-liquid-like
concept still holds for each region of the momentum and frequency
space), we analyze the imaginary part of the self-energy $\Sigma''\left(\mathbf{k},\omega\right)$
as a function of both frequency and temperature. In the top panels
of Fig.~\ref{fig7}, we plot the imaginary part of the self-energy
$\Sigma''\left(\underline{\mathbf{k}},\omega\right)$, together with
its real part $\Sigma'\left(\underline{\mathbf{k}},\omega\right)$,
the spectral function $A(\underline{\mathbf{k}},\omega)$ and the
non-interacting dispersion $\varepsilon_{0}\left(\underline{\mathbf{k}},\omega\right)$
as functions of the frequency at the nodal point $\underline{\mathbf{k}}=S$
(left panel) and at its \emph{companion} position on the \emph{phantom}
half of the pocket $\underline{\mathbf{k}}=\underline{S}$ (right
panel) along the main diagonal $\Gamma\to M$. In both cases, although
for $\underline{\mathbf{k}}=\underline{S}$ is not visible in the
picture, but is very clear for $\underline{\mathbf{k}}=S$, the position
of the relative/local maximum of $A(\underline{\mathbf{k}},\omega)$
coincides with the chemical potential (i.e. where are on the Fermi
surface) and it is determined by the sum of $\varepsilon_{0}\left(\underline{\mathbf{k}},\omega=0\right)$
and $\Sigma'\left(\underline{\mathbf{k}},\omega=0\right)$ as expected
(i.e. both points belong to the $r(\mathbf{k})=0$ locus). What is
very interesting and somewhat unexpected and peculiar, is that at
the nodal point a \emph{parabolic-like} (i.e. a Fermi-liquid-like)
behavior of $\Sigma''\left(\underline{\mathbf{k}},\omega\right)$
is clearly apparent, whereas at $\underline{\mathbf{k}}=\underline{S}$,
the dependence of $\Sigma''\left(\underline{\mathbf{k}},\omega\right)$
on frequency shows a predominance of a linear term giving a definite
proof that this region in momentum space is interested to a kind of
physics very different from what can be considered even by far Fermi-liquid
like. 

In order to remove any possible doubt regarding the non-Fermi-liquid-like
nature of the physics going on at the \emph{phantom} half of the pocket,
in the bottom panels of Fig.~\ref{fig7}, the imaginary part of the
self-energy at the Fermi surface $\Sigma''\left(\underline{\mathbf{k}},\omega=0\right)$
is reported as a function of the temperature at the nodal point $\underline{\mathbf{k}}=S$
and at its \emph{companion} $\underline{\mathbf{k}}=\underline{S}$.
The blue straight line in the right panel is just a guide to the eye.
We clearly see that in this case too, although it requires to move
from a $T$ to a $T^{2}$ representation (from left to right panel),
the behavior of $\Sigma''\left(\underline{\mathbf{k}},\omega=0\right)$
shows rather different behaviors at the selected points. In particular,
the temperature dependence of $\Sigma''\left(\underline{\mathbf{k}},\omega=0\right)$
is exactly parabolic (i.e., exactly Fermi-liquid) at the nodal point,
while it exhibits a predominance of linear and logarithmic contributions
at $\underline{S}$. This is one of the most relevant results of this
analysis and characterize this scenario with respect to the others
present in the literature.

\begin{figure}[p]
\noindent \centering{}\mbox{}\hspace*{\fill}\includegraphics[height=6cm]{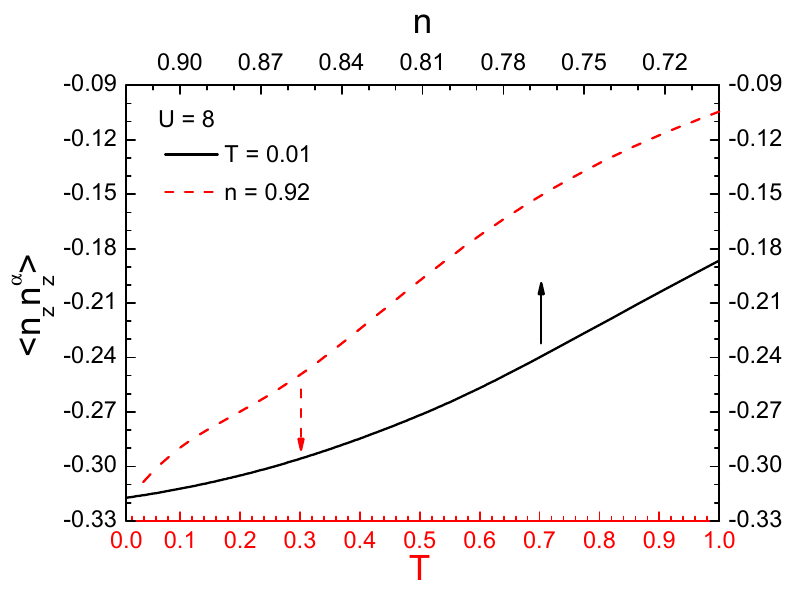}\hspace*{\fill}\includegraphics[height=6cm]{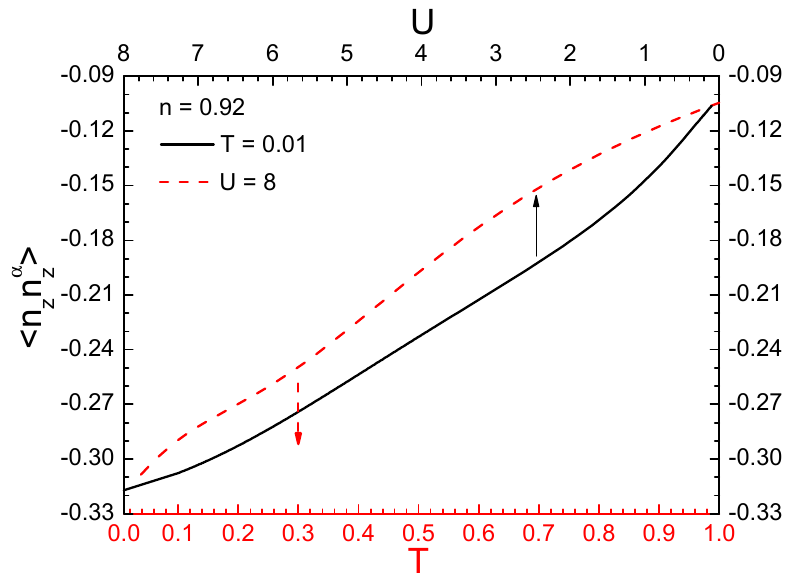}\hspace*{\fill}\mbox{}\protect\caption{(Color online) The spin-spin correlation function $\left\langle n_{z}n_{z}^{\alpha}\right\rangle $
as a function of filling $n$, temperature $T$ and on-site Coulomb
repulsion $U$ in the ranges $0.7<n<0.92$, $0.01<T<1$ and $0.1<U<8$.\label{fig11}}
\end{figure}

\begin{figure}[p]
\noindent \centering{}\mbox{}\hspace*{\fill}\includegraphics[height=6cm]{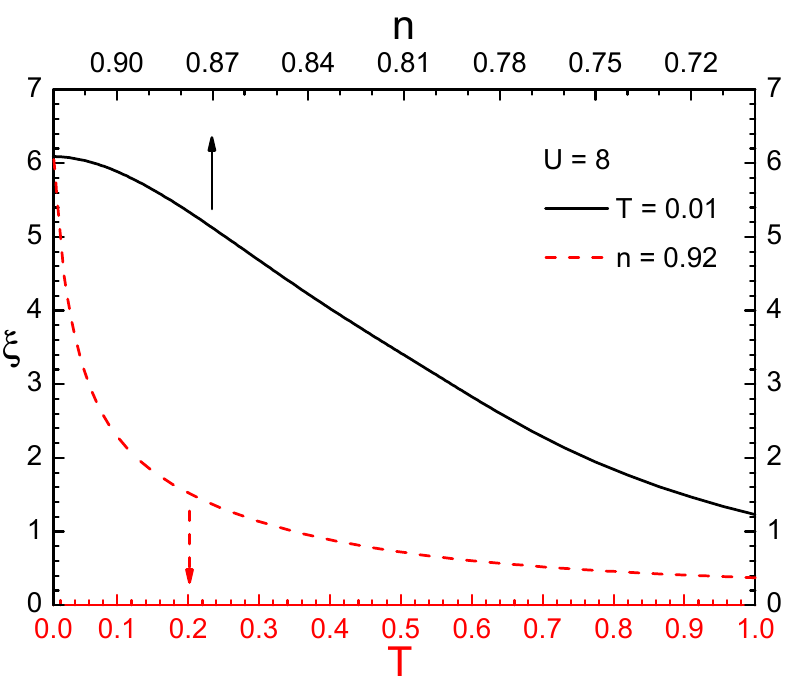}\hspace*{\fill}\includegraphics[height=6cm]{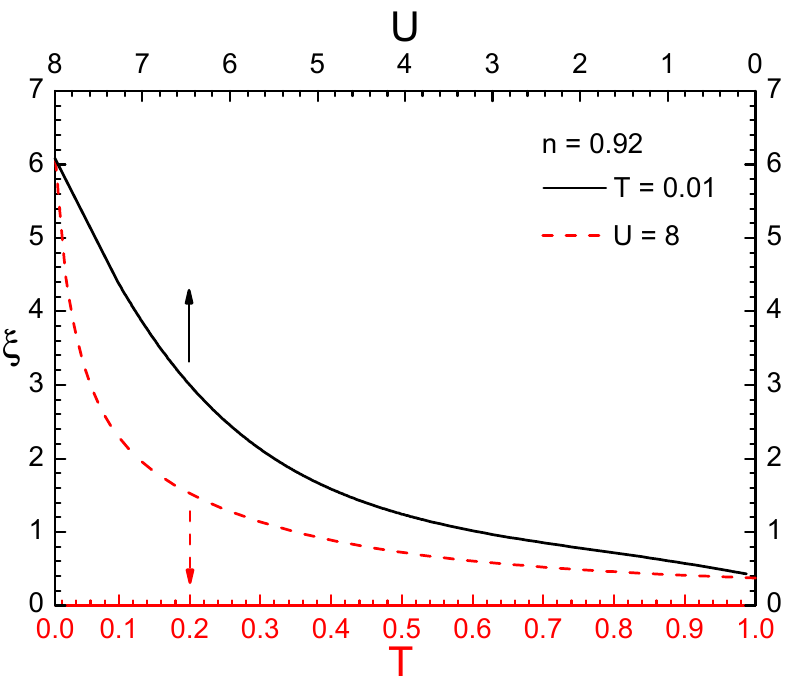}\hspace*{\fill}\mbox{}\protect\caption{(Color online) The antiferromagnetic correlation length $\xi$ as
a function of filling $n$, temperature $T$ and on-site Coulomb repulsion
$U$ in the ranges $0.7<n<0.92$, $0.01<T<1$ and $0.1<U<8$.\label{fig12}}
\end{figure}

\begin{figure}[p]
\noindent \centering{}\mbox{}\hspace*{\fill}\includegraphics[height=6cm]{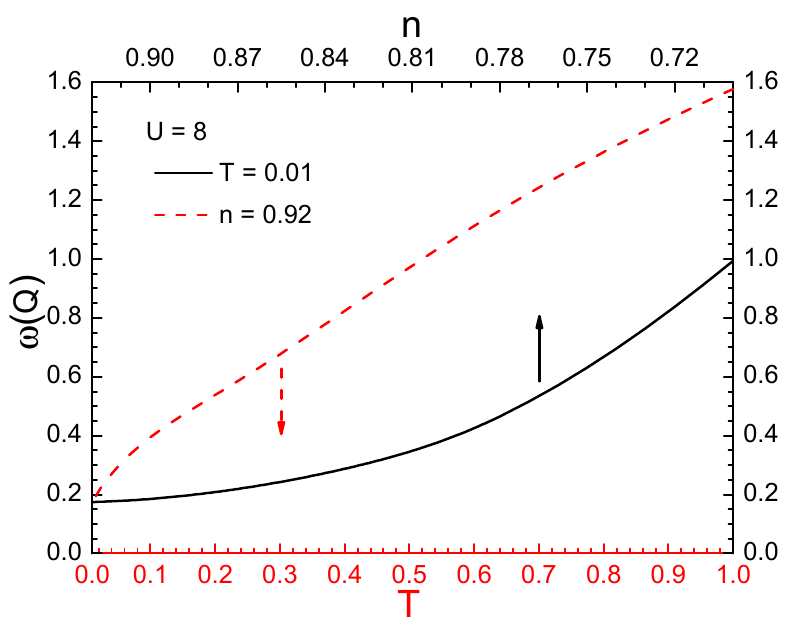}\hspace*{\fill}\includegraphics[height=6cm]{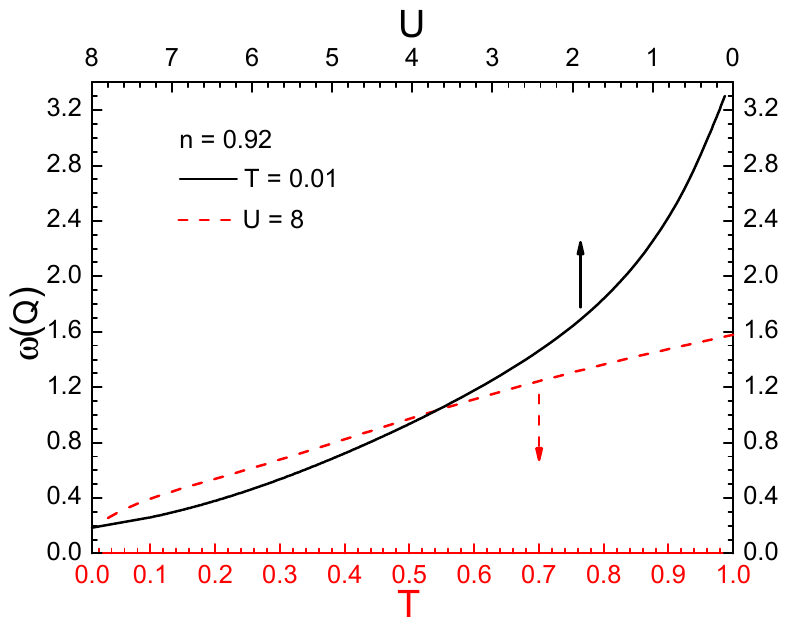}\hspace*{\fill}\mbox{}\protect\caption{(Color online) The pole $\omega^{(3)}(\mathbf{Q})$ of the spin-spin
propagator as a function of filling $n$, temperature $T$ and on-site
Coulomb repulsion $U$ in the ranges $0.7<n<0.92$, $0.01<T<1$ and
$0.1<U<8$.\label{fig13}}
\end{figure}

\subsection{Spin dynamics\label{sec:Spin-dynamics}}

Finally, to analyze the way the system approaches the antiferromagnetic
phase on decreasing doping and temperature and increasing correlation
strength, we report the behavior of the nearest neighbor spin-spin
correlation function $\left\langle n_{z}n_{z}^{\alpha}\right\rangle $,
of the antiferromagnetic correlation length $\xi$ and of the pole
$\omega^{(3)}(\mathbf{Q})$ (as defined in Sec.~\ref{sec:Results})
as functions of filling $n$, temperature $T$ and on-site Coulomb
repulsion $U$. In Figg.~\ref{fig11}, \ref{fig12} and \ref{fig13},
respectively, such quantities are presented for values in the ranges
$0.7<n<0.92$, $0.01<T<1$ and $0.1<U<8$. The choice for the extremal
values of low doping $n=0.92$, low temperature $T=0.01$ and strong
on-site Coulomb repulsion $U=8$ has been made as for these values
we find that all investigated single-particle properties (spectral
density function, Fermi surface, dispersion relation, density of states,
momentum distribution function, self-energy, ...) present anomalous
behaviors. The nearest-neighbor spin-spin correlation function $\left\langle n_{z}n_{z}^{\alpha}\right\rangle $
is always antiferromagnetic in character (i.e. negative) and increases
its absolute value on decreasing doping and temperature $T$ and on
increasing $U$ as expected. It is rather evident the signature of
the exchange scale of energy $J\approx\frac{4t^{2}}{U}\approx0.5$
in the temperature dependence as a significative enhancement in the
slope. The analysis of the filling dependence unveils quite strong
correlations at the higher value of doping too: $\xi$ is always larger
than one for all values of fillings showing that, at $T=0.01$ and
$U=8$, we should expect antiferromagnetic fluctuations in the overdoped
regime too as also claimed by recent experiments \cite{Le-Tacon_13,Dean_13,Jia_14}.
In the overdoped region, the antiferromagnetic fluctuations are quite
less well defined, in terms of magnon/paramagnon width, than in the
underdoped region \cite{Jaklic_95,Jaklic_95a,Vladimirov_09} and than
what found in the current two-pole approximation. In fact, a proper
description of the paramagnons dynamics would definitely require the
inclusion of a proper self-energy term in the charge and spin propagators
too (in preparation). Coming back to results (Fig.~\ref{fig12}),
$\xi$ overcomes one lattice constant at temperatures below $J$ and
tends to diverge for low enough temperatures. On the other hand, $\xi$
seems to saturate for low enough values of doping. $\xi$ equals one
between $U=3$ and $U=4$ and again rapidly increases for large enough
values of $U$. The pole $\omega^{(3)}(\mathbf{Q})$ decreases on
decreasing doping $n$ and temperature $T$ and on increasing $U$.
In particular, it is very sensitive to the variations in temperature
$T$ and in on-site Coulomb repulsion $U$, which make the mode softer
and softer clearly showing the definite tendency towards an antiferromagnetic
instability.

\section{Conclusions and Perspectives\label{sec:Conclusions}}

We have reviewed and systematized the theory and the results for the
single-particle and the magnetic-response properties microscopically
derived for the 2D Hubbard model, as minimal model for high-$T_{c}$
cuprate superconductors, within the Composite Operator Method with
the residual self-energy computed in the Non-Crossing Approximation
(NCA).

Among the several scenarios proposed for the pseudogap origin \cite{Norman_05},
COM definitely falls into the AF scenario (the pseudogap is a precursor
of the AF long-range order) \cite{Kampf_90,Schmalian_99,Abanov_03}
as well as the two-particle self-consistent approach (TPSC) \cite{Vilk_95,Tremblay_06},
the DMFT$+\Sigma$ approach \cite{Sadovskii_01,Sadovskii_05,Kuchinskii_05,Kuchinskii_06,Kuchinskii_06a}
and a Mori-like approach by Plakida and coworkers \cite{Plakida_01,Plakida_06,Plakida_10}.

In the limit of strong on-site Coulomb repulsion and low doping, such
results show the emergence of a pseudogap scenario, the deconstruction
of the Fermi surface in ill-defined open arcs and the clear signatures
of non-Fermi-liquid features similarly to what has been found by ARPES
experiments \cite{Damascelli_03} and not only \cite{Sebastian_12a}.
In particular, we have shown that a very low-intensity signal develops
around $M$ point and moves towards $S$ nodal point on decreasing
doping up to close, together with the ordinary Fermi surface boundary,
a pocket in the underdoped region. Whenever the pocket develops, it
is just the remarkable difference in the intensity of the signal between
the two halves of the pocket to make a Fermi arc apparent. As the
doping decreases further, the arc shrinks into a point at $S$ exactly
at half filling making possible to reconcile the large-small Fermi
surface dichotomy once the Fermi surface is defined as the relative
maxima of the spectral function and the relic of the ordinary paramagnetic
Fermi surface is also taken into account. The pseudogap develops since
a region in momentum (and frequency) with a very low-intensity signal
(it corresponds to the \emph{phantom} half of the pocket at the chemical
potential and to the shadow band out of it) is present between the
van Hove singularity and the quite flat band edge (quite flat after
the doubling of the Brillouin zone due to the very strong antiferromagnetic
fluctuations). On changing doping, a spectral weight transfer takes
place in the density of states between the two maxima corresponding
to the van Hove singularity and the band edge, respectively. A crossover
between a Fermi liquid and a non-Fermi liquid can be clearly observed
in the momentum distribution function (definitely not featuring a
sharp jump on the \emph{phantom} half of the pocket) and in the imaginary
part of the self-energy (featuring linear and logarithmic terms in
the frequency and temperature dependence instead of the ordinary parabolic
term) on decreasing doping at low temperatures and large interaction
strength. This crossover exactly corresponds to the process of deconstruction
of the Fermi surface. We also report kinks in the dispersion along
nodal and anti-nodal directions. In order to properly interpret the
behavior of the spectral density function and of the momentum distribution
function, we have also analyzed the characteristic features in the
spin-spin correlation function, the antiferromagnetic correlation
length and the pole of the spin-spin propagator. As expected, on reducing
doping or temperature and on increasing $U$, the correlations become
stronger and stronger. The exchange scale of energy $J$ is clearly
visible in the temperature dependence of the spin-spin correlation
function and drives the overall behavior of the magnetic response.
These results also demonstrate that a properly microscopically derived
susceptibility can give results practically identical or, at least,
very similar to those attainable by means of phenomenological susceptibilities
specially tailored to describe experiments. This is even more remarkable
since COM brings the benefice of a microscopical determination of
the temperature and filling dependencies of the correlation length. 

Many other issues should be addressed in the next future: to improve
the Hamiltonian description by adding and fine tuning longer-range
hopping terms in order to quantitatively and not only qualitatively
describe specific materials, to verify the stability and the modifications
of this scenario with respect to the inclusion of a residual self-energy
in the calculation of the charge and spin propagators closing a fully
self-consistent cycle, to investigate the charge and spin responses
in the full range of momentum and frequency relevant for these systems
searching for hourglasses, thresholds and all other peculiar features
experimentally observed, to investigate the superconducting phase
and establish it is nature and relationship with the anomalous features
of the \emph{normal} phases, to analyze in detail the transition/crossover
between the quasi-ordinary antiferromagnetic phase at half-filling
and the underdoped regime.
\begin{acknowledgments}
The author gratefully acknowledges many stimulating and enlightening
discussions with A. Chubukov, F. Mancini, N.M. Plakida, P. Prelovsek,
J. Tranquada and R. Zayer. He also wishes to thank E. Piegari for
her careful reading of the manuscript.
\end{acknowledgments}

\bibliographystyle{apsrev}
\bibliography{paper.bib}

\end{document}